\documentclass[aps,prd,11pt,superscriptaddress,notitlepage,longbibliography,nofootinbib,tightenlines]{revtex4-1}

\usepackage{amsmath,amssymb,amsfonts}
\usepackage{graphicx}
\usepackage{subfigure}
\usepackage{color}
\usepackage{hyperref}
\definecolor{darkred}{rgb}{0.8,0.1,0.1}
\hypersetup{colorlinks=true, linkcolor=darkred, citecolor=blue, linktoc=page}

\newcommand{\N}[1]{\ensuremath{\mathcal N=#1}}

\DeclareMathOperator{\diag}{diag}

\makeatletter\def\l@subsubsection#1#2{}%
\makeatother

\begin{document}

\title{Generalized gravitational entropy of probe branes: flavor~entanglement~holographically}

\author{Andreas Karch}
\email{akarch@u.washington.edu}
\author{Christoph F.~Uhlemann}
\email{uhlemann@uw.edu}
\affiliation{Department of Physics, University of Washington, Seattle, WA 98195-1560, USA}

\begin{abstract}
The notion of generalized gravitational entropy introduced by Lewkowycz and Maldacena 
allows, via the AdS/CFT correspondence, to calculate CFT entanglement entropies.
We adapt the method to the case where flavor branes are present and treated in the probe approximation.
This allows to calculate the leading flavor correction to the CFT entanglement entropy from the
on-shell action of the probe, while dealing with the backreaction is avoided entirely and from the outset.
As an application we give concise derivations for the contribution of massless and massive flavor
degrees of freedom to the entanglement entropy in \N{4} SYM theory.
\end{abstract}


\maketitle
\tableofcontents

\section{Introduction}
Entanglement entropy has received much attention in recent years, with applications ranging
from condensed matter systems to the holographic reconstruction of spacetime.
Given a physical system composed of two subsystems $A$ and $B$, it provides a measure 
of the extent to which information from one subsystem is relevant for the other.
For quantum field theories with a dual AdS/CFT description it can be calculated holographically, at least when
the subsystems arise from a spatial partition of the background geometry into regions $A$ and $B$.
The entanglement entropy of the region $A$ is then given by the area of 
a certain minimal surface extending to the boundary of AdS.
Namely, it should end there on the entangling surface, which is the boundary $\partial A$ of the region $A$ \cite{Ryu:2006bv}.

This rather ad-hoc proposal for the holographic calculation was placed on firmer ground in \cite{Lewkowycz:2013nqa}.
The key is to introduce a notion of gravitational entropy which extends the usual 
finite-temperature equilibrium interpretation of Euclidean compact-time solutions \cite{Gibbons:1976ue}
to the case where there is no U(1) isometry along the S$^1$ time direction.
The gravitational entropy of a configuration with a, possibly asymptotic, boundary which has an S$^1$ direction is calculated by 
considering a family of solutions, where the period of the S$^1$ is varied. 
More explicitly, 
with $S(n)$ denoting the on-shell gravity action for the solution with period $2\pi n$, 
the gravitational entropy $\mathcal S_\mathrm{g}$ is given by
\begin{align}\label{eqn:S_EE-Renyi}
 \mathcal S_\mathrm{g}&=-n\partial_n \left[\log Z(n)-n\log Z(1)\right]_{n\rightarrow 1}~,&\log Z(n) = -S(n)~.
\end{align}
The configuration itself is kept periodic with the original period $2\pi$ also for $n\neq 1$.
The non-trivial part then is to evaluate (\ref{eqn:S_EE-Renyi}) for the case where $n$ is not an integer and translations along the S$^1$ are not a symmetry:
there is a clash in keeping the original period for the boundary conditions and implementing the varying identifications along the S$^1$ direction. 
One of the two equivalent prescriptions given in \cite{Lewkowycz:2013nqa} is to avoid that problem by defining $S(n)$ as follows.
One integrates the S$^1$ direction only over $[0,2\pi)$ in the action.
Since the geometry should be regular for a period $2\pi n$, this introduces an apparent conical singularity 
with opening angle $2\pi/n$ at the place where the S$^1$ degenerates, already for the U($1$) symmetric case.
To get the full action the result is then multiplied by $n$, such that $S(n)=n S(n)_{2\pi}$. 
Using this prescription in (\ref{eqn:S_EE-Renyi}) yields
\begin{align}\label{eqn:S_EE-Renyi2}
 \mathcal S_\mathrm{g}&=\lim_{n\rightarrow 1}n^2\partial_n S(n)_{2\pi}~.
\end{align}

A connection to the holographic calculation of entanglement entropies arises for the case where the gravity solution is a Euclidean 
asymptotically-AdS space with a specific boundary geometry.
Namely, the S$^1$ direction should on the boundary encircle the entangling surface $\partial A$.
The family of bulk solutions labeled by $n$ then produces on the boundary $n$-fold covers of the original geometry,
branched along $\partial A$. 
These are precisely the geometries that would be used to calculate the entanglement entropy 
directly in the CFT, as a limit of Renyi entropies with the replica trick.
With the standard AdS/CFT identification of the bulk and boundary partition functions \cite{Maldacena:1997re,Witten:1998qj,Gubser:1998bc},
the calculation of the gravitational entropy (\ref{eqn:S_EE-Renyi}) is then equivalent to the calculation
of the entanglement entropy in the dual theory with the replica trick.
Moreover, as argued in \cite{Lewkowycz:2013nqa}, this formula reduces to the area of the minimal surface ending on $\partial A$, 
so it reproduces the proposal of \cite{Ryu:2006bv}.

A topic of recent interest are the entanglement entropy corrections arising when flavor degrees of freedom are 
added to the CFT, which have been studied, e.g., in \cite{Chang:2013mca,Jensen:2013lxa,Kontoudi:2013rla,Lewkowycz:2013laa}. 
Adding flavors in the quenched approximation corresponds in the bulk to the addition of branes in the probe approximation.
To calculate the leading-order contribution of the flavors to the entanglement entropy with the method of \cite{Ryu:2006bv},
one has to compute the backreaction of the flavor branes on the bulk geometry and then the resulting change in the area of the minimal surface.
As discussed in \cite{Chang:2013mca}, one can avoid an explicit calculation of the backreaction by expressing it as a convolution of
the brane energy momentum tensor with the gravitational Green's function. 
The resulting double-integral formula offers a crucial simplification: it turns out that the 
detailed properties of the internal space only mildly affect the calculation and can be subsumed into an effective brane energy-momentum tensor.
Nevertheless, that one needs the backreaction at all may seem surprising, given that the leading corrections to
other quantities, like the thermal entropy density, can be calculated from the on-shell action of the probe alone.
In this paper we show that the method of \cite{Lewkowycz:2013nqa} offers a new perspective on that issue:
by a suitable adaption we can get the leading-order correction to the entropy without computing the backreaction.
The calculation of the entanglement entropy from on-shell actions, combined with extremality arguments similar to those used in \cite{Lewkowycz:2013nqa},
allows us to argue that the brane embeddings are only needed for the $n=1$ geometry, and that we do not need the backreaction altogether.
This naturally is not entirely for free. With the branes treated in the probe approximation we can not generally reduce the 
expression (\ref{eqn:S_EE-Renyi2}) to a pure boundary term, and the computation does not explicitly reduce to the area of the 
Ryu/Takayanagi minimal surface.
However, this can certainly be a reasonable trade for avoiding the backreaction or a double integral of twice the dimension.

As a specific application we consider the D3/D7 setup \cite{Karch:2002sh}, which yields a holographic description of \N{4} super Yang-Mills theory 
coupled to flavor degrees of freedom, and the D3/D5 system which adds flavors confined to a codimension-$1$ hypersurface to
the Yang-Mills theory \cite{Karch:2000gx,DeWolfe:2001pq}.
We calculate the entanglement entropy of a spherical region $A$.
For the pure CFT that entropy has been derived in \cite{Casini:2011kv}, by conformally mapping it to a thermal one.
To calculate the entanglement entropy from the gravitational entropy (\ref{eqn:S_EE-Renyi2}), we need the one-parameter 
family of bulk solutions described above. 
For $n\neq 1$ these turn out to be the Euclidean versions of hyperbolic AdS black holes,
and to get the flavor contribution we have to consider the D5 and D7 branes in these backgrounds.
Nicely enough, though, it is still sufficient to know the extremal brane embedding for the $n\,{=}\,1$ AdS geometry.
The entanglement entropies for massless flavors have previously been calculated in \cite{Jensen:2013lxa,Chang:2013mca}, 
and our method reproduces those results.
We also consider the case where the D7 branes are separated from the stack of D3 branes, which corresponds to adding massive flavors to the CFT.
Building on the backreacted D3/D7 bulk geometry obtained by means of a smearing technique in \cite{Bigazzi:2009bk}, their contribution to 
the entanglement entropy has been studied holographically in \cite{Kontoudi:2013rla}.
In our approach we will not have to deal with the complicated backreaction, and the case provides an example where the
brane embedding breaks the U(1) isometry along the S$^1$ direction.
We find that the universal terms in the entanglement entropy agree in both calculations.
Comparing the remaining scheme-dependent terms is difficult, since keeping the regularization scheme unaffected by the 
flavor perturbation is subtle in the backreaction approach.
To allow for a more detailed comparison we also calculate the change in the Ryu/Takayanagi minimal area with the double integral formula 
derived in \cite{Chang:2013mca}. This simplifies the backreaction approach rather drastically and allows us to better keep track of the
regularization procedure.
Up to differences in the finite terms, which as we will argue should be expected, we then find completely agreeing results.

The paper is organized as follows. 
We start in Sec.~\ref{sec:GGE-bulk} by calculating the pure CFT entanglement entropy of a spherical region in terms of the 
gravitational entropy of the dual gravity theory, and discuss in detail the relevant family of bulk solutions. 
In Sec.~\ref{sec:GGE-probe-branes} we give a general discussion of how branes are incorporated into the
generalized gravitational entropy method and how the probe approximation can be exploited very efficiently.
This method is then applied in Sec.~\ref{sec:EE-from-GGE} to calculate the entanglement entropy corrections due to 
massless and massive flavors in \N{4} SYM theory.
In Sec.~\ref{sec:double-integral} we give an independent calculation of the massive case, 
following the double-integral approach of \cite{Chang:2013mca}, and compare to the previous results.
We conclude in the final Sec.~\ref{sec:conclusion}.

\section{Entanglement entropy from generalized gravitational entropy}\label{sec:GGE-bulk}

To set the stage and fix notation we calculate the entanglement entropy of a spherical region in pure \N{4} SYM.
The result itself has already been derived in \cite{Ryu:2006bv,Casini:2011kv,Jensen:2013lxa,Guo:2013jva},
but the calculation provides a nice opportunity to highlight the facts about the gravitational entropy calculation that 
will be relevant for the probe brane discussion.
We consider the Euclidean CFT on flat $\mathbb{R}^{d}$ with line element
\begin{align}
 ds^2=dt^2+dr^2+r^2d\Omega_{d-2}^2~.
\end{align}
The region $A$ for which we want to calculate the entanglement entropy corresponds to $r<\ell$ at $t=0$,
and we denote the complement at $t=0$ by $B$.
If we wanted to calculate the entanglement entropy directly as limit of Renyi entropies in the boundary theory
by the replica trick, we would consider $n$-fold covers of the background geometry, branched along $\partial A$.
Such geometries can be obtained by the coordinate transformation
\begin{align}\label{eqn:boundary-change-of-coordinates}
 t&=\frac{\ell\sin\tau}{\cosh u+\cos\tau}~,&  r&=\frac{\ell\sinh u}{\cosh u+\cos \tau}~.
\end{align}
This covers the entire $\mathbb{R}^d$ and maps the regions $A$ and $B$ to  
$A=\lbrace \tau=0, u\in\mathbb{R}\rbrace$ and $B=\lbrace\tau=\pi,u\in\mathbb{R}\rbrace$, respectively.
The line element becomes
\begin{align}\label{eqn:n-fold-boundary-metric}
 ds^2&=\Omega^2\left(d\tau^2+du^2+\sinh^2(u)d\Omega_{d-2}^2\right)~,\qquad \Omega=\ell(\cosh u+\cos\tau)^{-1}~.
\end{align}
The period of the S$^1$ direction $\tau$ naturally is $2\pi$, and adjusting the range to $0\leq\tau<2\pi n$ yields the desired $n$-fold covers.

To calculate the CFT entanglement entropy holographically from the generalized gravitational entropy of the dual gravity theory, 
we have to consider asymptotically AdS$_5\times$S$^5$ solutions which yield on the boundary of AdS$_5$ the geometry (\ref{eqn:n-fold-boundary-metric}), 
with the S$^1$ direction encircling $\partial A$.
That is, the boundary geometry is precisely the $n$-fold cover that would be used in the replica trick. 
The generalized gravitational entropy of the $n\,{=}\,1$ solution can then be calculated from (\ref{eqn:S_EE-Renyi2})
and yields the CFT entanglement entropy.

\subsection{Bulk geometry for a spherical entangling surface}\label{sec:geometry}
\begin{figure}[ht]
\center
\subfigure[][]{ \label{fig:AdS-S1H3-1}
  \includegraphics[width=0.2\linewidth]{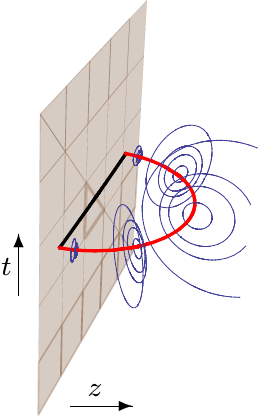}
}\qquad\qquad\qquad\qquad
\subfigure[][]{ \label{fig:AdS-S1H3-2}
    \includegraphics[width=0.23\linewidth]{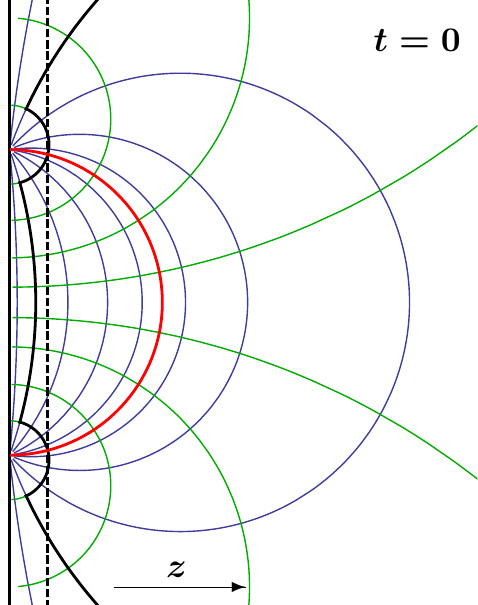}
}
\caption{Illustration of how the coordinates (\ref{eqn:coord-transf}) cover the Poincar\'{e} patch of Euclidean AdS.\label{fig:AdS-S1H3}
         The plane on the left hand side represents the boundary at $z=0$, the black thick line is $A$, the red thick line is the
         place where the S$^1$ direction $\tau$ degenerates, and the blue curves are along constant $\rho$ and $u$.
         The right hand side shows a section of the figure on the left hand side through the plane $t=0$.
         The blue circular lines correspond to constant $\rho$ and the green perpendicular ones to constant $u$.
         At the red half circle corresponding to $\rho=\ell^{-1}$ the S$^1$ direction degenerates. 
         The part inside of it has $\tau=0$ and the outside part $\tau=\pi$.}
\end{figure}

We now discuss the AdS bulk geometry such that the boundary takes the form (\ref{eqn:n-fold-boundary-metric}).
Starting with AdS in Poincar\'{e} coordinates, $ds^2=L^2z^{-2}\left(dz^2+dt^2+dr^2+r^2d\Omega^2\right)$,
we can extend the coordinate transformation (\ref{eqn:boundary-change-of-coordinates}) into the bulk and perform the transformation
\begin{align}\label{eqn:coord-transf}
 z&=\frac{1}{\rho\cosh u+\tilde\rho\cos\tau}~,&
 t&=z \ell\tilde\rho\sin\tau~,&
 r&=z \ell\rho\sinh u~,
\end{align}
where $\tilde\rho^2=\rho^2-\ell^{-2}$. For $\rho\rightarrow\infty$ this turns into (\ref{eqn:boundary-change-of-coordinates}).
The resulting geometry is the S$^1\times \mathbb{H}^{d-1}$ slicing of AdS$_{d+1}$ with line element
\begin{align}\label{eqn:AdS-S1H3}
 ds^2&=L^2\Big(\frac{d\rho^2}{\rho^2-\ell^{-2}}+(\rho^2\ell^2-1)d\tau^2+\rho^2 \ell^2 ds_{\mathbb{H}^{d-1}}^2\Big)~,&
 ds_{\mathbb{H}^{d-1}}^2&=du^2+\sinh^2(u)d\Omega_{d-2}^2~,
\end{align}
where the $\mathbb{H}^{d-1}$ slices have radius of curvature $\ell$.
Demanding that there be no conical singularity at $\rho=\ell^{-1}$ yields the identification $\tau\sim\tau+2\pi$.
The Lorentzian version of that geometry covers the causal completion of the spherical region $A$ on the boundary of AdS, 
as discussed in \cite{Casini:2011kv}.
The Euclidean version, however, covers all of the Poincar\'{e} patch, as illustrated in Fig.~\ref{fig:AdS-S1H3}.
This figure already shows that the place where the S$^1$ degenerates corresponds to the minimal area ending on $\partial A$, which was
used to calculate the CFT entanglement entropy in \cite{Ryu:2006bv}.

The boundary metric which we extract from that bulk geometry depends on the choice of defining function $h$: 
following the standard procedure \cite{Graham:1999jg}, the metric on the conformal boundary is defined by 
\begin{align}
 g_\mathrm{bndy}:=\lim_{\epsilon\rightarrow 0}h^2 g\vert^{}_{\partial \mathcal M_\epsilon}~,
\end{align}
where $\mathcal M_\epsilon$ denotes the asymptotically-AdS spacetime with a finite spatial cut-off, 
and $\partial\mathcal M_\epsilon$ the resulting boundary in the ordinary sense.
If we now take the bulk metric (\ref{eqn:AdS-S1H3}) and choose $h=1/(\rho L)$, 
the resulting boundary geometry is S$^1\times \mathbb{H}^{d-1}$.
On the other hand, keeping the defining function we had originally used in Poincar\'{e} coordinates,
$h=z(\rho,\tau,u)/L$ with $z$ as given in (\ref{eqn:coord-transf}), 
we get precisely (\ref{eqn:n-fold-boundary-metric}) as boundary geometry.
Changing the defining function corresponds to a conformal transformation in the boundary theory,
and for a CFT the choice is up to us.
Since, however, we will be interested also in the case where massive flavors are added to the CFT, 
such that conformal invariance is broken by a relevant deformation, 
we have to keep the latter defining function used in Poincar\'{e} coordinates,
and thus get (\ref{eqn:n-fold-boundary-metric}) on the boundary.

It is now not too hard to find the bulk solutions for $n\,{\neq}\,1$: these are 
the Euclidean versions of the hyperbolic AdS black holes discussed in \cite{Emparan:1999gf}.
With $C_h=\rho_h^d-\rho_h^{d-2}\ell^{-2}$, the line element reads
\begin{align}\label{eqn:n-fold-bulk-metric}
 ds^2&=L^2\left(\frac{d\rho^2}{f_n(\rho)}+f_n(\rho)\ell^2 d\tau^2+\rho^2\ell^2ds_{\mathbb{H}^{d-1}}^2\right)~,& 
 f_n(\rho)&=\rho^2-\ell^{-2}-C_h\rho^{2-d}~.
\end{align}
The position of the horizon is determined from the absence of a conical singularity for $\tau\sim\tau+2\pi n$.
This yields $\ell f_n^\prime(\rho_h)=2/n$, or more explicitly 
\begin{align}
 \ell^2\rho_h^2 d-(d-2)= 2\rho_h\ell/n~.
\end{align}
Translations along $\tau$ are an isometry of the metrics for all $n$, and for $n=1$ the geometry reduces to~(\ref{eqn:AdS-S1H3}).
The above discussion of the extraction of the boundary geometry and the choices of defining functions
also applies for $n\neq 1$, since $f_n(\rho)$ becomes independent of $n$ for large $\rho$.
The boundary geometry thus is -- depending on the defining function -- either S$^1\times \mathbb{H}^{d-1}$ or (\ref{eqn:n-fold-boundary-metric}), 
but with period $2\pi n$.
So, with the choice of defining function discussed above, we have indeed found the family of solutions needed to evaluate the entanglement
entropy from (\ref{eqn:S_EE-Renyi2}).

To simplify the following computations, we note that we can also just work with the bulk geometry (\ref{eqn:n-fold-bulk-metric}) with $\ell=1$.
To get the boundary metric (\ref{eqn:n-fold-boundary-metric}) we then have to use the defining function $h\,{=}\,z(\rho,\tau,u) \ell/L$.
This change of defining function corresponds to a constant rescaling of the metric on the field-theory side. 
The usual identification of the field theory UV cut-off $\epsilon^{-1}$ with the bulk IR cut-off $z>\epsilon$
is thus changed to $z>\epsilon/\ell$. 
Likewise, the bulk quantity corresponding to the mass $M_q$ of the flavor fields in the CFT 
(up to a rescaling the separation of the flavor branes from the D3 branes) now corresponds
to the dimensionless quantity $M_q\ell$ on the CFT side.
This alternative way of encoding the radius $\ell$ of $A$ in the bulk computation will be used in Sec.~\ref{sec:EE-from-GGE}.

\subsection{Gravitational entropy of pure gravity}\label{sec:nobranes}
We now calculate the gravitational entropy of the solution (\ref{eqn:AdS-S1H3}) via (\ref{eqn:S_EE-Renyi2}), 
which yields the entanglement entropy for the spherical region $A$ in the dual CFT without flavors.
To get the on-shell action for Einstein-Hilbert gravity on the bulk geometries (\ref{eqn:n-fold-bulk-metric}) we have to employ the usual
procedure of holographic renormalization. 
That is, after cutting off the bulk spacetime at a finite spatial distance, we supplement the action by covariant counterterms
on the cut-off surface, to cancel the divergences.
Here we have to deal with two kinds of divergences:  those arising for large AdS$_{d+1}$ radial coordinate $\rho$ and those for large
$\mathbb{H}^{d-1}$ radial coordinate $u$, and we introduce a cut-off in both directions,
analogously to the procedure in \cite{Andrade:2011nh}.
The variation of the finite renormalized actions with $n$ would then automatically produce a finite entropy.
However, in the existing calculations of the entanglemenet entropy the divergences of the minimal area are kept, to see, 
e.g., the QFT area law \cite{Srednicki:1993im} and to have the universal $\log$ terms accessible.
As we want to compare to the existing results, 
we implement the following partial renormalization of the bulk action: we subtract off the large-$\rho$ divergences by
adding the usual holographic counterterms at a cut-off surface of fixed large $\rho=\rho_\epsilon$. 
This leaves divergences arising for large $u$, which we do not cancel. 
Following \cite{Casini:2011kv}, we fix the cut-off by noting that, at the horizon
$\rho=\ell^{-1}$, we have from (\ref{eqn:coord-transf}) that $z=\ell/\cosh u$.
With the usual identification of the field theory cut-off with the bulk IR cut-off, we thus introduce an upper bound on $u$,
which is given by
\begin{align}\label{eqn:cut-off-u}
 \cosh u_\mathrm{max}&=\ell/\epsilon~.
\end{align}
We then have the cut-off surface shown in Fig.~\ref{fig:AdS-S1H3-2}, but introduce counterterms
only on the part of constant $\rho=\rho_\epsilon$, which excludes the circular regions\footnote{%
One may be worried that this could add counterterms at the Poincar\'{e} horizon,
corresponding here to $\tau\,{=}\,\pi$, $u\,{=}\,0$, $\rho\,{\rightarrow}\,\infty$.
This can be avoided by a lower cut-off on $u$, in addition to (\ref{eqn:cut-off-u}), such that the surface
where counterterms are added does not cross a fixed $z=\eta$.
This yields $\cosh u\,{>}\,(1-\eta\tilde\rho_\epsilon\cos\tau)/(\eta\rho_\epsilon)$, which,
as $\rho_\epsilon\rightarrow\infty$, only removes the point $z\rightarrow\infty$ from the cut-off surface.
However, none of our calculations is sensitive to this procedure.}.

We can now turn to the actual calculation.
The S$^5$ part of the on-shell action just contributes a factor of the volume, 
which we absorb into the definition of the Newton constant, $G:=G_{10d}/V_{S^5}$.
We will also keep the dependence on $d$ for AdS$_{d+1}$ explicit, since the final result generalizes accordingly,
with an appropriate definition of $G$ to account for different internal spaces.
After integrating over the internal space, we use the (partly) renormalized action
$S_\mathrm{bulk,ren}=S_\mathrm{bulk}+S_\mathrm{ct}$ with
\begin{subequations}\label{eqn:bulk-gravity-action}
\begin{align}
 S_\mathrm{bulk}&=-\frac{1}{16\pi G}\int_{\mathcal M}d^{d+1}x\sqrt{g}\Big(R[g]+\frac{d(d-1)}{L^2}\Big)~, \label{eqn:effective-5d-gravity-action}\\
 S_\mathrm{ct}&=-\frac{1}{16\pi G}\int_{\partial \mathcal M}d^dx\sqrt{g_\epsilon}\Big(2K-\frac{2(d-1)}{L}+\frac{L}{d-2}R[g_\mathrm{\epsilon}]+\dots\Big)~.
\end{align}
\end{subequations}
The metric (\ref{eqn:n-fold-bulk-metric}) is a solution to the bulk field equations, where we use curvature conventions such that
the curvature of AdS$_{d+1}$ is $R=-d(d+1)/L^2$.
In the second line, $\partial \mathcal M$ denotes the boundary at a fixed large $\rho_\epsilon$ and $g_\epsilon$ the metric induced there. 
There potentially are more counterterms, depending on $d$, and specifically for $d=4$ we have the usual $\log$-divergent and finite counterterms 
constructed from the squared Weyl tensor of $g_\epsilon$.
However, the induced metric on the cut-off surface here is S$^1\times\mathbb{H}^{d-1}$ and thus conformally flat, such that these terms vanish.
For the metric determinants and the extrinsic curvature, $K=\frac{1}{2}g_\mathrm{ind}^{\mu\nu}\mathcal L_N g_{\mathrm{ind}\,\mu\nu}$,
we have
\begin{align}\label{eqn:bulk-determinants-etc}
 \sqrt{g}&=L^{d+1}\ell^d\rho^{d-1}\sqrt{g_{\mathbb{H}^{d-1}}}~,&
 \sqrt{g_\epsilon}&=L^d\ell^df^\frac{1}{2}\rho^{d-1}\sqrt{g_{\mathbb{H}^{d-1}}}~,&
 K&=\frac{d-1}{\rho L}f^\frac{1}{2}+L^{-1}\partial_\rho f^\frac{1}{2}~.
\end{align}
Since the induced metric on $\partial\mathcal M$ is $S^1\times\mathbb{H}^{d-1}$,
the scalar curvature is just the sum of the $S^1$ and $\mathbb{H}^{d-1}$ curvatures, $R[g_\epsilon]=-\rho_\epsilon^{-2}L^{-2}(d-1)(d-2)/\ell^2$.

With all the ingredients at hand we can now calculate the gravitational entropy of the $n\,{=}\,1$ 
solution (\ref{eqn:AdS-S1H3}) by evaluating (\ref{eqn:S_EE-Renyi2}), which yields the entanglement entropy $\mathcal S_\mathrm{EE}$ in the dual CFT. 
The $n$-dependent on-shell action is obtained by just integrating $\tau$ over $[0,2\pi)$, such that
\begin{align}\label{eqn:S_EE-bulk-general}
 \mathcal S_\mathrm{EE}&=n^2\partial_n \left[S_\mathrm{bulk}(n)_{2\pi}+S_\mathrm{ct}(n)_{2\pi}\right]_{n=1}~.
\end{align}
A striking point of \cite{Lewkowycz:2013nqa} is that the computation of the gravitational entropy actually reduces to the
evaluation of boundary terms.
The arguments used there rely on rewriting the bulk part such that it is proportional to the equation of motion, which produces
additional boundary terms at $\rho_\epsilon$ and $\rho_h$.
For the explicit example at hand we can see that rather straightforwardly: 
the only $n$-dependence in the integrands of (\ref{eqn:bulk-gravity-action})
is through $f_n(\rho)$, and the integrand of $S_\mathrm{bulk}$ evaluates to $-2dL^{-2}\sqrt{g}$, which is actually independent of 
$f_n$ and $n$.
Without using any integration by parts, only 
taking into account that the lower bound of the $\rho$-integration, $\rho_h=\rho_h(n)$, depends on $n$, the variation with $n$ thus reduces to
\begin{align}
 -16\pi GL^{2} \,n^2\partial_n S_\mathrm{bulk} \ &= \ n^2\partial_n \int_{\rho_h}^{\rho_\epsilon}d\rho d^dx(-2d)\sqrt{g}
 \ = \ -(n^2\partial_n\rho_h)\int_{\rho=\rho_h} d^dx(-2d)\sqrt{g}~.
\end{align}
There is no contribution from a change in the range of the $\tau$-integral, since that was restricted to $[0,2\pi)$ for all $n$.
Both contributions in (\ref{eqn:S_EE-bulk-general}) are thus reduced to boundary terms. 
In fact, the $\tau$ integrations are trivial here, since translations along $\tau$ are an isometry, 
so both terms calculate -- up to coefficients -- the volume of $\mathbb{H}^{d-1}$, which is 
also the minimal surface ending on $\partial A$.
For the evaluation of the counterterms we note that the
derivative of $f_n$ with respect to $n$ is strongly suppressed for large $\rho$, namely
\begin{align}\label{eqn:delta_n-f-suppression}
 \partial_n f_n(\rho)&=\mathcal O(\rho^{-d}f_n(\rho))~.
\end{align}
Thus, only the GHY term and the volume counterterm, which are the leading terms at large $\rho_\epsilon$
and of $\mathcal O(\rho_\epsilon^d)$, can contribute finite parts to the gravitational entropy.
With the explicit expressions given in (\ref{eqn:bulk-determinants-etc}) above and
\begin{align}
 \partial_n f_n(\rho)\vert_{n=1}&=\frac{2\rho^{2-d}}{\ell^{d}(d-1)}~,& \partial_n\rho_h\vert_{n=1}=\frac{1}{\ell(1-d)}~,
\end{align}
we can then evaluate (\ref{eqn:S_EE-bulk-general}).
With $V_{\mathbb{H}^{d-1}}$ denoting the (regularized) volume of $\mathbb{H}^{d-1}$ with unit radius of curvature,
this yields
\begin{align}\label{eqn:SEE-gen}
 \mathcal S_\mathrm{EE}&=\frac{1}{4G}L^{d-1}V_{\mathbb{H}^{d-1}}~.
\end{align}
The AdS$_{d+1}$ minimal surface ending on the sphere $\partial A$ on the boundary is precisely
$\mathbb{H}^{d-1}$ (as we will see explicitly in Sec.~\ref{sec:double-integral-EE} below), so 
this result agrees with the previous derivations.
Evaluating the volume of $\mathbb{H}^{d-1}$ with the cut-off $u\leq u_\mathrm{max}$ as given in (\ref{eqn:cut-off-u}), we find
\begin{align}\label{eqn:bulk-gravity-EE-explicit}
 \mathcal S_\mathrm{EE}&=\frac{L^{d-1}V_{\mathrm{S}^{d-2}}}{4G}\int_0^{u_\mathrm{max}}du\sinh^{d-2}u
 =\frac{L^{d-1}V_{\mathrm{S}^{d-2}}}{4G}
  \int_{\epsilon/\ell}^1 ds\frac{(1-s^2)^{(d-3)/2}}{s^{d-1}}~.
\end{align}
The explicit form of the volume of S$^{d-2}$ is $V_{S^{d-2}}=2\pi^{(d-1)/2}/\Gamma(\frac{d-1}{2})$ and to
get the second equality we have substituted $\cosh(u)=1/s$. 
This result precisely reproduces (3.3) of \cite{Ryu:2006bv}, where $d$ refers to AdS$_{d+2}$.
We also see explicitly that we could have carried out the computation with the $\ell=1$ geometry, just taking
into account the modified identification of the gravity and field theory cut-offs, and we will follow that
procedure from now on.
Expanding (\ref{eqn:bulk-gravity-EE-explicit}) for small $\epsilon$ yields
\begin{align}\label{eqn:bulk-gravity-EE-expanded}
 \mathcal S_\mathrm{EE}&=\frac{L^{d-1}V_{\mathrm{S}^{d-2}}}{4G}\left(\frac{\ell^2}{2 \epsilon ^2}+\frac{1}{2}\log \frac{\epsilon}{2\ell}-\frac{1}{4}+O\left(\epsilon ^1\right)\right)~.
\end{align}
The divergent and finite parts generally depend on the regularization scheme and on the choice of state in the CFT,
from which the entanglement entropy is calculated, while the universal, scheme-independent information is in the coefficient 
of the $\log$-term \cite{Ryu:2006ef}.

\section{Gravitational entropy of probe branes: Reduction to boundary terms vs.\ probe approximation}\label{sec:GGE-probe-branes}
We now include branes embedded into the bulk spacetime into the picture. 
The gravitational entropy of the combined system of bulk gravity and embedded brane, described by a 
total action $S=S_\mathrm{bulk}+S_\mathrm{brane}$, can again be calculated via (\ref{eqn:S_EE-Renyi2}).
Since solving the combined system of equations for the bulk fields and brane embedding functions is 
notoriously difficult, the branes are often treated in the probe approximation.
This corresponds to solving for the brane embedding in a fixed gravitational background which is a solution to the bulk 
equations of motion.
The backreaction of the brane on the bulk geometry is then only taken into account perturbatively.
Since the background solution, into which the brane is embedded, extremizes the bulk action, the bulk part has an expansion of
the form 
\begin{align}\label{eqn:Sbulk-t0-expansion}
S_\mathrm{bulk}=S_\mathrm{bulk}^{(0)}+t_0^2S_\mathrm{bulk}^{(2)}+\dots~, 
\end{align}
where $t_0$ is the parameter controlling the strength of the backreaction (a combination of Newton's constant and the brane tension).
The salient feature is, of course, the absence of a term linear in $t_0$.
The brane action, on the other hand, itself is of $\mathcal O(t_0)$ and including the change of the embedding due to the linearized backreaction
again only produces terms of $\mathcal O(t_0^2)$.
For quantities calculated directly from the on-shell action, the probe approximation therefore yields correct results 
at linear order in $t_0$ even without including the backreaction at all.
This unfortunately does not apply to the area of minimal surfaces embedded into the bulk spacetime: their area is certainly sensitive to 
$\mathcal O(t_0)$ corrections to the bulk metric. The extremality of the bulk action just does not (obviously) help here.
For a holographic calculation of the flavor contribution to the entanglement entropy via \cite{Ryu:2006bv}, even at leading order, one thus 
has to actually calculate the backreaction.
In the following we will discuss the gravitational entropy contribution of the branes. 
As it turns out, here we can obtain the results to $\mathcal O(t_0)$ without calculating the backreaction.
Moreover, it is also enough to know the correct brane embedding, determined from extremizing the DBI action, just for the $n=1$ geometry.

We consider an on-shell bulk configuration with a boundary which has an S$^1$ direction, possibly without U(1) isometry along the S$^1$,
into which a brane should be embedded.
The gravitational entropy can then once again be calculated with the formula (\ref{eqn:S_EE-Renyi2}), which becomes
\begin{align}\label{eqn:S-GGE-bulk-brane}
 \mathcal S_\mathrm{g}&=n^2\partial_n \left[S_\mathrm{bulk}(n)_{2\pi}+S_\mathrm{brane}(n)_{2\pi}\right]_{n=1}~.
\end{align}
We are instructed to find a family of bulk configurations where the S$^1$ direction is covered 
$n$ times, with $n$ possibly non-integer. 
Restricting the range of integration for the S$^1$ direction to $[0,2\pi)$ introduces an apparent conical singularity with opening
angle $2\pi/n$, whether or not the configuration has a U($1$) isometry.
For the following general arguments we use a notation similar to \cite{Lewkowycz:2013nqa}, where $\tau$ denotes the coordinate along 
the S$^1$ direction and $r$ the direction along which the S$^1$ degenerates. 
The locus where the S$^1$ degenerates is $r\,{=}\,0$, and the boundary corresponds to $r\,{\rightarrow}\,\infty$.
Since we will be explicitly interested in asymptotically-AdS spaces, where both the bulk and brane actions have to be renormalized
by introducing a cut-off and counterterms at large $r$, we will explicitly include the counterterms into the arguments.
For other configurations one just has to appropriately drop or replace the counterterm contribution.
We then have
\begin{align}
 S_\mathrm{bulk}(n)&=\int dr\, d^dx  L_\mathrm{bulk} \ + \ S_\mathrm{ct,bulk}~,&
 S_\mathrm{brane}(n)&=\int dr\, d^py  L_\mathrm{brane} \ + \ S_\mathrm{ct,brane}~,
\end{align}
where $x$, $y$ denote the transverse coordinates in the bulk and on the brane, respectively.
We start with the contribution of the bulk action to the entropy.
The derivative with respect to $n$ simply becomes
\begin{align}
 n^2\partial_n S_\mathrm{bulk}(n)_{2\pi}\big\vert_{n=1}&=\int_0^{r_\epsilon} dr\, d^dx\, n^2\partial_n L_\mathrm{bulk}
 \ + \ n^2\partial_n S_\mathrm{ct,bulk}\Big\vert_{n=1}~.
\end{align}
The derivative with respect to $n$ evaluated at $n=1$ can be understood as a first-order variation, which we simply write as $\delta_n$.
Following \cite{Lewkowycz:2013nqa}, we can now rewrite the derivative of the bulk Lagrangian using integration by parts, 
such that the integrand becomes proportional to the equations of motion.
We use the notation 
\begin{align}
 \delta_n L&= \frac{\delta L}{\delta g_{\mu\nu}}\delta_n g_{\mu\nu} + \partial_\mu\Theta^\mu[\delta_n g]~,
\end{align}
to separate the part yielding the equations of motion from the total derivatives, and analogously for other fields
and the brane action.
The total derivatives potentially produce boundary terms at the cut-off at large $r_\epsilon$, as well as at the apparent conical singularity at $r=0$.
The former are then combined with the variation of the counterterm action, and we find
\begin{align}\label{eqn:deltan-Sbulk}
\begin{split}
 \delta_n S_\mathrm{bulk}(n)_{2\pi}&=
 \int_0^{r_\epsilon} dr\, d^dx\, \frac{\delta L_\mathrm{bulk}}{\delta g_{\mu\nu}}\delta_n g_{\mu\nu}\\&\hphantom{=}
 +\int_{r=r_\epsilon}d^dx\, \big(\delta_n L_\mathrm{ct,bulk}+N^\epsilon_\mu \Theta_\mathrm{bulk}^\mu[\delta_n g]\big)
 +\int_{r\sim 0}d^dx\,N^0_\mu\Theta_\mathrm{bulk}^\mu[\delta_n g]~,
\end{split}
\end{align}
where the metric represents the entire set of bulk fields.
We have also denoted the outward-pointing unit normal vector fields to the surfaces $r=r_\epsilon$ and $r\sim 0$
by $N^\epsilon$ and $N^0$, respectively.
If we had the bulk theory alone, we would
conclude that for an on-shell configuration the first term vanishes.
The second term vanishes as well, since the holographic counterterms are constructed precisely such that
the renormalized action is stationary for solutions of the bulk field equations satisfying appropriate boundary conditions.
We would thus be left with the third term, which evaluates to the area of the locus where the S$^1$ degenerates, as argued
in Sec.~3.2 of \cite{Lewkowycz:2013nqa}.

We now want to add probe branes to the system.
For simplicity we will also assume that the brane Lagrangian only involves the volume of the induced metric, the generalization
to additional and possibly higher-derivative brane fields should be straightforward.
In general, the  brane embedding will adjust to changes in $n$, and one has to solve the entire system of resulting equations for each $n$.
The variation of the brane action then becomes
\begin{align}
\begin{split}
 \delta_n S_\mathrm{brane}&=\int drd^py\left(
 \frac{\delta L_\mathrm{brane}}{\delta g_{\mu\nu}}\delta_n g_{\mu\nu}+\frac{\delta L_\mathrm{brane}}{\delta X_\mu}\delta_n X^\mu\right)\\&\hphantom{=}
 +\int_{r=r_\epsilon}d^py \left(\delta_n L_\mathrm{ct,brane}+N^\epsilon_\mu\Theta_\mathrm{brane}^\mu[\delta X]\right)
 +\int_{r\sim 0} d^py N^0_\mu\Theta_\mathrm{brane}^\mu[\delta X]~.
\end{split}
\end{align}
There are no boundary terms from the variation with respect to the bulk metric, since the Lagrangian does not involve its derivatives.
The combination of boundary terms at $r_\epsilon$ vanishes when evaluated for on-shell embedding functions $X^\mu$, since the
brane counterterms are again constructed such that the action is stationary.
Moreover, in contrast to the analogous terms for the bulk theory, the boundary terms produced at $r\sim 0$ vanish as well:
in contrast to the curvature at a conical singularity, the volume stays perfectly finite and shrinks to zero as the tip of the cone is approached.
Thus, as long as the brane Lagrangian just involves the volume form, there is no additional contribution from $r\sim 0$.
This may change as curvature terms are included in the effective brane action, and a
nice discussion of curvature invariants at conical singularities can be found in \cite{Fursaev:1995ef,Fursaev:2013fta}.
As emphasized in \cite{Lewkowycz:2013nqa}, one should also not add explicit boundary terms at $r=0$, since the conical singularity
is just an apparent one, resulting from our restriction to $\tau\in[0,2\pi)$, while the full geometries are regular.
If we assume the embedding functions $X^\mu$ to be on shell for $n\,{=}\,1$, we are thus left with
\begin{align}\label{eqn:deltan-Sbrane}
 \delta_n S_\mathrm{brane}&=\int drd^py
 \frac{\delta L_\mathrm{brane}}{\delta g_{\mu\nu}}\delta_n g_{\mu\nu}
 \ + \ \int_{r=r_\epsilon}d^py \frac{\delta  L_\mathrm{ct,brane}}{\delta g_{\mu\nu}}  \delta_n g_{\mu\nu}
 ~.
\end{align}
The contribution from the brane counterterms is now reduced to the variation with respect to changes in the bulk metric,
which enters through the volume and curvatures of the induced metric on the cut-off surface on the brane.

We could now use the backreacted metric to evaluate (\ref{eqn:S-GGE-bulk-brane}) with (\ref{eqn:deltan-Sbulk}) and (\ref{eqn:deltan-Sbrane}). 
The metric variations in (\ref{eqn:deltan-Sbrane}) then cancel the first two  terms in (\ref{eqn:deltan-Sbulk}).
This is just the statement that the backreacted metric solves the combined Einstein equations, and that the counterterms are 
constructed to cancel the boundary terms at $r_\epsilon$.
We are then left with the third term in (\ref{eqn:deltan-Sbulk}), evaluated for the backreacted metric.
This of course just yields the area of the minimal surface, now calculated for the backreacted metric, 
and thus reproduces the prescription of \cite{Ryu:2006bv}.

But the derivation this far also allows us to take a different route and avoid calculating the backreaction:
Similarly to the arguments given around (\ref{eqn:Sbulk-t0-expansion}) above, we can exploit the probe approximation
to get the entropy to linear order in $t_0$.
Since the bulk geometries for all $n$ are constructed such that they extremize the bulk action
(we are deforming along an on-shell path), we can again use the extremality argument
and expand, for each $n$,
\begin{align}
 S_\mathrm{bulk}(n)&=S_\mathrm{bulk}^{(0)}(n)+\mathcal O(t_0^2)~,
\end{align}
where $S_\mathrm{bulk}^{(0)}(n)$ is evaluated on the non-backreacted metric.
The probe brane contribution is itself $\mathcal O(t_0)$, and including the backreaction there also only produces $\mathcal O(t_0^2)$ corrections.
We can thus use the non-backreacted metrics to evaluate (\ref{eqn:S-GGE-bulk-brane}) to linear order in $t_0$.
This yields the first two terms in the expansion $\mathcal S_\mathrm{g}=\mathcal S_\mathrm{g}^{(0)}+\mathcal S_\mathrm{g}^{(1)}+\mathcal O(t_0^2)$,
where $t_0$ is implicit in $\mathcal S_\mathrm{g}^{(1)}$.
When evaluated on the non-backreacted metric, the zeroth-order term (\ref{eqn:deltan-Sbulk}) again reduces to the boundary term at the 
locus where the S$^1$ degenerates:
the first two terms vanish since the bulk equations of motion are solved and the third yields the minimal area.
The leading-order correction (\ref{eqn:deltan-Sbrane}) now can not be further reduced to a boundary term, which means that we will 
actually have to calculate the integrated variation of the brane Lagrangian.
Summing up, we find for the gravitational entropy to linear order in $t_0$
\begin{align}\label{eqn:S-GGE-simplified}
 \mathcal S_g&=\frac{A_\mathrm{min}[g]}{4G}+\int drd^py
 \frac{\delta L_\mathrm{brane}}{\delta g_{\mu\nu}}\delta_n g_{\mu\nu}
 \ + \ \int_{r=r_\epsilon}d^py \frac{\delta  L_\mathrm{ct,brane}}{\delta g_{\mu\nu}}  \delta_n g_{\mu\nu}~.
\end{align}
The first term is $\mathcal S_\mathrm{g}^{(0)}$ and arises from the last term in (\ref{eqn:deltan-Sbulk}), 
while the remaining terms are linear in $t_0$ and yield $\mathcal S_\mathrm{g}^{(1)}$.
The bottom line of this general discussion thus is, that we can calculate the contribution of the probe branes to the gravitational
entropy by just considering the variation of the DBI action with respect to changes in the bulk metric with $n$, as given in (\ref{eqn:S-GGE-simplified}).
We do not need to calculate the backreaction and we only need the correct brane embedding for the $n\,{=}\,1$ geometry.

\subsection{Euclidean hyperbolic AdS black holes}
We now put the calculation of Sec.~\ref{sec:nobranes} in perspective and evaluate (\ref{eqn:S-GGE-simplified})
for the bulk geometry~(\ref{eqn:n-fold-bulk-metric}).
This will then be used in the next section to calculate the flavor contribution to
the CFT entanglement entropy of the spherical region $A$.
In the coordinates used in Sec.~\ref{sec:nobranes}, changing $n$ also changes the range of the radial coordinate $\rho$,
which we have to take into account. 
For the variation of the bulk action away from $n\,{=}\,1$, which produces the first term of 
$\mathcal S_\mathrm{EE}=\mathcal S_\mathrm{EE}^{(0)}+\mathcal S_\mathrm{EE}^{(1)}+\mathcal O(t_0^2)$, we found
\begin{align}\label{eqn:bulk-variation}
 \mathcal S_\mathrm{EE}^{(0)}&=\delta_n S_\mathrm{bulk}(n)_{2\pi}=-\delta\rho_h \int_{\rho=\rho_h(1)}d^dxL_\mathrm{bulk}
 +\int_{\rho>\rho_h(1)}d\rho\, d^dx\, \delta_n L_\mathrm{bulk}+\delta_n S_\mathrm{ct,bulk}~,
\end{align}
where $\delta\rho_h=n^2\partial_n\rho_h(n)\vert_{n=1}$. 
Since the $\tau$ integral is fixed to $[0,2\pi)$, there was no contribution from a change of the $\tau$ interval.
As we had seen in Sec.~\ref{sec:nobranes}, the on-shell Lagrangian was independent of $n$ already, so rather than
using integration by parts to rewrite it as equation of motion plus boundary terms, we decided to explicitly 
evaluate the expression at this point already, to get $\mathcal S_\mathrm{EE}^{(0)}$.
For the brane embedded into (\ref{eqn:n-fold-bulk-metric}) we analogously find
\begin{align}\label{eqn:S-EE-probe-brane}
 \mathcal S_\mathrm{EE}^{(1)}&=\delta_n S_\mathrm{brane}(n)_{2\pi}=
 -\delta\rho_h \int\limits_{\rho=\rho_h}d^pyL_\mathrm{brane}
 +\int\limits_{\rho_h}^{\rho_\epsilon}d\rho d^py \frac{\delta L_\mathrm{brane}}{\delta g_{\mu\nu}}\delta_n g_{\mu\nu}+\delta_n S_\mathrm{ct,brane}~.
\end{align}
As argued above, we have to take into account only the variation with respect to the bulk metric.
If we further assume the brane Lagrangian to be given solely by the volume form, without contributions of, e.g., worldvolume gauge fields,
the DBI action for a D$p$-brane just is 
\begin{align}
 L_\mathrm{brane}=-T_p\sqrt{\gamma}~,
\end{align}
with the induced metric $\gamma$ on the worldvolume of the D$p$-brane.
We note that a positive-tension brane has negative $T_p$ with that sign convention.
To calculate the contribution to the entanglement entropy we need to evaluate (\ref{eqn:S-EE-probe-brane}),
which then becomes
\begin{align}\label{eqn:S-EE-D7}
 \mathcal S^{(1)}_\mathrm{EE}&=
  \delta\rho_h\,T_p\int_{\rho=\rho_h} d^py\sqrt{\gamma}
 \ - \ T_p\int_{\rho_h}^{\rho_\epsilon}d\rho\int d^py\, n^2\partial_n\sqrt{\gamma}
 \ + \ n^2\partial_nS_{\mathrm{ct,brane}}~.
\end{align}
Once again, we keep the $n=1$ brane embedding and only take into account the change
of the worldvolume due to the change in the spacetime metric.
This equation will be the starting point for the calculations of Sec.~\ref{sec:EE-from-GGE}.

\section{Flavor entanglement entropy from  generalized gravitational entropy}\label{sec:EE-from-GGE}
We now want to apply the method of the last section to include the contribution of probe branes to the gravitational entropy.
This will yield the entanglement entropy of the dual CFT with flavor degrees of freedom, which, corresponding to the
probe approximation, are treated in the quenched approximation.
Specifically, we consider the D3/D5 and D3/D7 systems, which we introduce in the following. 
From the spacetime perspective the D7 branes provide an example of spacetime-filling branes, while the D5 branes are
codimension $1$.
The mass of the flavors in the dual theory depends on the 
separation of the flavor branes from the D3 branes in the ten-dimensional spacetime.
We take the AdS$_5\times$S$^5$ background created by the D3 branes in coordinates where the AdS$_5$ part is given
by (\ref{eqn:AdS-S1H3}) and the S$^5$ part reads
\begin{align}\label{eqn:metric-S5}
 ds^2_{\text{S}^5}&=L^2\left(d\psi^2+\cos^2(\psi) d\theta^2+\sin^2(\psi) d\Omega_3^2\right)~,
 & d\Omega_3^2&=d\varphi^2+\sin^2(\varphi)d\Omega_2^2~.
\end{align}
The brane embeddings extremizing the DBI action can simply be obtained by transforming the solutions
given in \cite{Karch:2002sh} to our coordinates.
The D7 branes wrap an AdS$_5\times$S$^3$ subspace of the D3-brane near-horizon geometry, which is defined by $\theta=0$ and
\begin{align}\label{eqn:D7-embedding}
 \cos\psi=\mu z=\frac{\mu}{\rho\cosh u+\tilde \rho \cos\tau}~,
\end{align}
where $\mu$ is proportional to the separation between the D7 and D3 branes.
The separation of the branes corresponds to the flavor mass in units of the string tension, 
and with the usual AdS/CFT identifications this yields a quark mass $M_q=\frac{\sqrt{\lambda}}{2\pi}\mu$,
as given, e.g., in \cite{Herzog:2006gh}.
However, we will use the rescaling discussed in Sec.~\ref{sec:geometry} to set $\ell\,{=}\,1$ in the metric (\ref{eqn:AdS-S1H3}),
which changes the identification of the bulk quantity $\mu$ with the flavor mass to 
\begin{align}
 M_q&=\frac{\sqrt{\lambda}}{2\pi}\frac{\mu}{\ell}~.
\end{align}
At $z=1/\mu$ the S$^3$ wrapped inside the S$^5$ vanishes and so does the effective tension seen from AdS$_5$,
such that the branes end there.
Choosing $\mu\neq 0$ breaks the isometries of AdS$_5$. 
However, in the hyperbolic slicing the amount of manifest symmetry is reduced rather drastically:
with $\tau$ and $u$ appearing in (\ref{eqn:D7-embedding}), the U(1) symmetry of translations along $\tau$ is broken, 
and only an SO($d-1$) remains of the $\mathbb{H}^{d-1}$ symmetries.
The ISO($d$) transformations acting on the slices of constant $z$ in Poincar\'{e} coordinates, 
or constant $z(\rho,\tau,u)$ in the hyperbolic slicing,
are of course symmetries in both coordinates, they may just not be as obvious.
The U(1) isometry along the S$^1$ direction $\tau$, however, is broken.
For the embedding of the D5 branes we make one of the angular variables of the hyperbolic slices in (\ref{eqn:AdS-S1H3}) explicit and write
\begin{align}
 ds_{\mathbb{H}^{d-1}}^2&=du^2+\sinh^2(u)\left(d\phi^2+\sin^2(\phi)d\Omega_{d-3}^2\right)~.
\end{align}
The D5 branes then wrap an AdS$_4\times$S$^2$ defined by $\phi=\varphi=\frac{\pi}{2}$, $\theta=0$ and (\ref{eqn:D7-embedding}).
Correspondingly, the flavor degrees of freedom in the dual theory are confined to a codimension-$1$ subspace.

In the following we calculate the gravitational entropy contribution of the D5 and D7 branes in the probe approximation.
To validate our resulting entanglement entropies in the massive case, we compare to the double-integral formula derived in \cite{Chang:2013mca},
which represents the change in the Ryu/Takayanagi minimal area due to the linearized backreaction.

\subsection{Massless flavors from D3/D7 and D3/D5}
We start with the massless case, $\mu=0$, when the D5/D7 probe branes are not separated from the D3 branes,
before turning to the more involved massive case.
As seen in Sec.~\ref{sec:nobranes}, we can in fact obtain rather general results and thus again keep the dependence on $d$ 
for AdS$_{d+1}$ explicit.
The brane embeddings in the massless case respect the U($1$) isometry of the background geometry in the $\tau$ direction.
They also preserve the SO($1,d\,{-}\,1$) symmetry of the $\mathbb{H}^{d-1}$ hyperbolic slices of AdS for the D7 and, correspondingly, 
the SO($1,d\,{-}\,2$) symmetry of the $\mathbb{H}^{d-2}$ slices for the D5 branes\footnote{%
For the D7 branes the $n\,{=}\,1$ embedding in fact extremizes the brane action also for $n\,{\neq}\, 1$,
since they wrap a maximal S$^3$ in S$^5$ and the entire AdS factor, which is the only part that changes with $n$.}.
We will use the bulk geometry with $\ell=1$ and, as explained above, incorporate the radius of the sphere
by a proper identification of bulk and boundary quantities. 
To keep the expressions simple we will also fix the AdS$_5$ and S$^5$ radii of curvature to $L=1$.

\subsubsection{D7 branes in \texorpdfstring{AdS$_5\times$S$^5$}{AdS5xS5}}

We now specialize to the D7 brane. 
To evaluate (\ref{eqn:S-EE-D7}) we just need the induced metric, which simply is the AdS$_5\times$S$^3$ part of the bulk metric
with line element
\begin{align}\label{eqn:D7-massless-induced-metric}
 ds^2_\gamma=\frac{d\rho^2}{f_n(\rho)}+f_n(\rho)d\tau^2+\rho^2 ds_{\mathbb{H}^{d-1}}^2+d\Omega_3^2~.
\end{align}
Nicely enough, $f_n(\rho)$ drops out of the induced volume form, which thus is actually independent of $n$. 
The second term in (\ref{eqn:S-EE-D7}) therefore vanishes in the massless case.
The S$^3$ part of the DBI action just contributes a factor of the volume, and with $\rho_h'(n)\vert_{n=1}=1/(1-d)$
the contribution from the D7-brane action to (\ref{eqn:S-EE-D7}) evaluates to
\begin{align}\label{eqn:D7-massless-horizon-deltan}
\rho_h^\prime(n)T_7\int_{\rho=\rho_h} d^7y\sqrt{\gamma}&=-\frac{1}{d-1}T_0V_{\Sigma}~,
\end{align}
where $T_0=T_7 V_{S^3}$ and  $V_\Sigma$ denotes the volume of the S$^1\times\mathbb{H}^{d-1}$ transverse to $\rho$.
This leaves the contribution of the counterterms to be evaluated. 
As shown in \cite{Karch:2005ms}, the usual covariant counterterms can be reorganized into 
those for just the AdS$_{d+1}$ part and those for a scalar field corresponding to the slipping mode $\psi$.
The relation of $\Phi$ used there to $\psi$ as given in (\ref{eqn:D7-embedding}) simply is $\Phi=\pi/2-\psi$.
The integral over the S$^3$ internal part then just produces a factor V$_{S^3}$, which turns $T_7$ into $T_0$.
For the massless case, $\psi=\pi/2$, the resulting counterterms are
\begin{align}\label{eqn:D7-ct-massless}
 S_\mathrm{D7,ct}&=T_0\int_{\rho=\rho_\epsilon}d^dy_s\sqrt{\gamma_{s,\epsilon}}\left(\frac{1}{d}-\frac{1}{2d(d-1)(d-2)}R[\gamma_{s,\epsilon}]\right)~,
\end{align}
where $\gamma_{s,\epsilon}$ denotes the metric induced from the AdS$_{d+1}$ part of the bulk metric on the cut-off surface.
We have given the coefficients for general $d$, noting that in higher dimensions additional counterterms will be required.
The $\log$-term for $d=4$ once again vanishes since the cut-off metric is conformally flat, and 
the sign difference to \cite{Karch:2005ms} is attributed to the fact that AdS has negative curvature in the conventions used here.
To evaluate the contribution of these counterterms to the entanglement entropy via (\ref{eqn:S-EE-D7}), 
we are interested in their derivative with respect to $n$. 
Due to the strong suppression of the derivative of $f_n$ with respect to $n$ by (\ref{eqn:delta_n-f-suppression}),
again only the volume counterterm can produce a non-vanishing contribution.
Evaluating the third term in (\ref{eqn:S-EE-D7}), we thus find
\begin{align}\label{eqn:D7-massless-ct-delta_n}
 n^2\partial_nS_{\mathrm{D7,ct}}&=\frac{1}{d(d-1)}T_0 V_\Sigma~.
\end{align}
One may be worried that, since the bulk geometry changes with $n$, the implementation of the field theory cut-off in the bulk theory also
depends on $n$. However, the asymptotic expansion of the bulk geometry does not change with $n$ up to terms which are suppressed by at least $\rho^{-d}$.
The interpretation of the bulk cut-off in the dual theory is thus only changed at a correspondingly subleading order, which does not
affect the entanglement entropy where the leading divergence is $\mathcal O(\epsilon^{2-d})$. 
Combining (\ref{eqn:D7-massless-horizon-deltan}) and (\ref{eqn:D7-massless-ct-delta_n}) with (\ref{eqn:S-EE-D7}), 
using $V_\Sigma=2\pi V_{\mathbb{H}^{d-1}}$, we thus find
\begin{align}\label{eqn:S_EE-D7massless}
 \mathcal S_\mathrm{EE}^{(1)}&=-\frac{\pi}{2}T_0V_{\mathbb{H}^{d-1}}=
 -\frac{t_0}{2d}\cdot \frac{V_{\mathbb{H}^{d-1}}}{4 G}~,
\end{align}
where we have introduced $t_0:=16\pi G T_0$ in the second equality.
This is precisely the result of (\ref{eqn:SEE-gen}) with an overall factor $-t_0/(2d)$,
and thus reproduces the result found in \cite{Chang:2013mca,Jensen:2013lxa}.

\subsubsection{D5 branes in \texorpdfstring{AdS$_5\times$S$^5$}{AdS5xS5}}
We now turn to the D5 brane. The induced metric on the brane is simply the AdS$_d\times$S$^2$ part
of the bulk geometry, with line element
\begin{align}
 ds^2_\gamma=\frac{d\rho^2}{f_n(\rho)}+f_n(\rho)d\tau^2+\rho^2 ds_{\mathbb{H}^{d-2}}^2+d\Omega_2^2~.
\end{align}
As in the D7 case, the induced volume form on the branes is independent of $n$ and (\ref{eqn:S-EE-D7})
reduces to the contribution from $\rho=\rho_h$ and from the counterterms. 
The former evaluates to
\begin{align}
\rho_h^\prime(n)T_5\int_{\rho=\rho_h} d^5y\sqrt{\gamma}&=-\frac{1}{d-1}T_5V_{S^2}V_{\Sigma}~.
\end{align}
$V_\Sigma$ now denotes the volume of $S^1\times\mathbb{H}^{d-2}$, but since we still have $\rho_h'(n)\vert_{n=1}=1/(1-d)$, the overall coefficient is not changed.
Turning to the counterterm contribution, we note that the leading counterterm, which is proportional to the volume of the cut-off slice,
only diverges as $\rho_\epsilon^{d-1}$. 
Due to (\ref{eqn:delta_n-f-suppression}), the counterterms therefore do not yield a contribution that survives the limit 
$\rho_\epsilon\rightarrow \infty$, in contrast to the D7 case.
The final result with  $T_0=T_5V_{S^2}$ thus reads
\begin{align}\label{eqn:S_EE-D5massless}
 \mathcal S_\mathrm{EE}^{(1)}&=-\frac{2\pi}{d-1}T_0V_{\mathbb{H}^{d-2}}=
 -\frac{t_0}{2(d-1)}\cdot \frac{V_{\mathbb{H}^{d-2}}}{4 G}~.
\end{align}
This again nicely reproduces the entropy corrections derived in \cite{Chang:2013mca,Jensen:2013lxa}.
We thus find that our gravitational-entropy results (\ref{eqn:S_EE-D7massless}), (\ref{eqn:S_EE-D5massless}), which required 
neither conformally mapping the entanglement entropy to a thermal one, 
nor computing the backreaction, agree with the existing results.

\subsection{Massive flavors from separated D3/D7}
We now turn to the case where D7 branes are separated from the D3 branes.
Of the AdS$_5\times$S$^5$ near-horizon geometry of the D3 branes they then
wrap the part of S$^5$ given by (\ref{eqn:D7-embedding}) with $\mu\neq 0$.
At $z=\mu^{-1}$ the D7 branes end in smoke, and to keep the focus on the essential
steps we will discuss the case of small mass, $\mu\,{<}\,1$.
In this case the branes cover the entire locus $\rho\,{=}\,1$ where the S$^1$ degenerates, 
which is not the case for large mass and has to be taken into account there.
Conformal invariance of the boundary theory is broken by the presence of the massive flavors,
which in the bulk is reflected by the breaking of the radial isometries evident in Poincar\'{e} coordinates.
A comment is in order on the change from Poincar\'{e} coordinates to the S$^1\times\mathbb{H}^{d-1}$ slicing (\ref{eqn:AdS-S1H3}).
Changing coordinates clearly is a perfectly valid thing to do, and since the S$^1\times\mathbb{H}^{d-1}$ slicing of AdS$_{d+1}$ 
covers the entire Poincar\'{e} patch in the Euclidean setting, we can actually perform all our calculations in these coordinates.
However, in the massless case, where conformal invariance was intact, we could have changed in addition the defining function to $h=1/(\rho L)$, 
which would have corresponded to considering the dual CFT on S$^1\times\mathbb{H}^{d-1}$.
Once conformal invariance is broken, switching to S$^1\times\mathbb{H}^{d-1}$ on the CFT side is not a symmetry anymore, and we 
have to do an honest change of coordinates, keeping the original defining function $h=z(\rho,\tau,u)/L$.

To evaluate (\ref{eqn:S-EE-D7}), we first need the induced metric $\gamma$ on the brane, for which we find
\begin{align}
 ds_\gamma^2&=\frac{d\rho^2}{f_n(\rho)}+f_n(\rho)d\tau^2+\rho^2(du^2+\sinh^2(u)d\Omega_2^2)+\frac{\mu^2dz^2}{1-\mu^2z^2}+(1-\mu^2z^2)d\Omega_3^2~.
\end{align}
To keep the expression simple, we have denoted $z(\rho,\tau,u)$ as given in (\ref{eqn:coord-transf}) simply by $z$, 
and analogously $dz=(\partial_\rho z) d\rho+(\partial_\tau z) d\tau+(\partial_u z) du$.
For $\mu=0$ this reduces to (\ref{eqn:D7-massless-induced-metric}).
We then need the determinant of the induced metric for $n=1$ and its derivative with respect to $n$ at $n=1$.
The former can easily be evaluated by transforming the result from Poincar\'{e} coordinates, where the induced metric is
given below in (\ref{eqn:D7inducedmetricPoincare}) and, 
with $dx^\mu dx_\mu=dt^2+dr^2+r^2d\Omega_2^2$, we have $\sqrt{\gamma}\vert_{n=1}=z^{-5}r^2\sqrt{g_{S^2\times S^3}} (1-\mu^2z^2)$. 
For the latter we have to actually evaluate the derivative of the determinant, to arrive at
\begin{align}\label{eqn:determinants}
 \sqrt{\gamma}_{n=1}&=\rho^3\sinh^2(u) \sqrt{g_{S^2\times S^3}}(1-\mu^2z^2)~,&
 \left[\partial_n\sqrt{\gamma}\right]_{n=1}&=\mu^2\sqrt{\gamma}_{n=1}\frac{(\tilde\rho^2\partial_\rho z)^2-(\partial_\tau z)^2}{3\rho^{2}\tilde\rho^{4}}~.
\end{align}
We start with the first term of (\ref{eqn:S-EE-D7}), i.e.\ the contribution from the boundary at the horizon.
Using (\ref{eqn:determinants}) with $\rho=1$ it evaluates to
\begin{align}\label{eqn:deltan-D7-massive-horizon-part}
\rho_h^\prime(n)T_7\int_{\rho=\rho_h} d^7y\sqrt{\gamma}&=-\frac{1}{3}T_7V_{S^3}\left(V_{\Sigma}-2\pi \mu^2V_{S^2}\int_0^{u_\mathrm{max}} du \tanh^2u\right)~.
\end{align}
Since we have assumed small mass, $\mu<1$, the restriction to $z<\mu^{-1}$ does not restrict the range of the $u$ integration.

We now turn to the counterterm contribution. 
In addition to the counterterms in (\ref{eqn:D7-ct-massless}) we now have those involving the slipping mode.
As explained above, the counterterms can be split into those constructed from the spacetime part of 
the brane metric for $\mu=0$, which is just AdS$_5$, and those involving $\psi$. 
In addition to the terms in (\ref{eqn:D7-ct-massless}) we now have 
\begin{align}\label{eqn:D7-massive-ct}
 S_\mathrm{D7,ct}^\psi=T_0\int_{\rho=\rho_\epsilon}d^4y_s\sqrt{\gamma_{s,\epsilon}}
 \Big(&\frac{1}{2}\Phi^2-\frac{1}{2}\log(\rho_\epsilon)\Phi\square^W_{\gamma_{s,\epsilon}}\Phi
 +\alpha\Phi^4+\beta\Phi\square^W_{\gamma_{s,\epsilon}}\Phi\Big)~,
\end{align}
where $\Phi=\pi/2-\psi=\arcsin \mu z(\rho,\tau,u)$ and 
$\square^W_{\gamma_{s,\epsilon}}=\square_{\gamma_{s,\epsilon}}-\frac{1}{6}R[\gamma_{s,\epsilon}]$
is the Weyl-covariant Laplacian\footnote{%
To cancel the large-$\rho_\epsilon$ divergence of the on-shell action without having to use integration by parts
we use a slightly modified form where 
$\Phi\square_{\gamma_{s,\epsilon}}\Phi\rightarrow 2\Phi\square_{\gamma_{s,\epsilon}}\Phi -\gamma_{s,\epsilon}^{\mu\nu}\partial_\mu\Phi\partial_\nu\Phi$.}.
The coefficients of the finite terms were fixed in \cite{Karch:2005ms} by demanding the on-shell
action to vanish, as required by supersymmetry.
One could in principle introduce an explicit $n$-dependence of the renormalization scheme by varying them with $n$.
However, this would introduce additional, spurious divergences, as these locally finite terms are integrated over
an infinite volume, and we thus keep them fixed.
The leading divergence in the counterterms (\ref{eqn:D7-massive-ct}) is just $\mathcal O(\rho_{\epsilon}^2)$, so due to 
(\ref{eqn:delta_n-f-suppression}) the derivatives with respect to $n$ vanish as $\rho_\epsilon\rightarrow\infty$.
This just leaves us with the contribution from the previously present counterterms (\ref{eqn:D7-ct-massless}), 
as given in (\ref{eqn:D7-massless-ct-delta_n}), and we find
\begin{align}\label{eqn:deltan-Sct-massive}
 n^2\partial_n(S_{\mathrm{D7,ct}}+S^\psi_{\mathrm{D7,ct}})&=\frac{t_0}{96G} V_{\mathbb{H}^3}~.
\end{align}
Using the explicit results for the counterterm variation, (\ref{eqn:deltan-Sct-massive}), 
and the contribution from the change in $\rho_h$, (\ref{eqn:deltan-D7-massive-horizon-part}), 
in (\ref{eqn:S-EE-D7}) yields
\begin{align}\label{eqn:S-EE-massive1}
 \mathcal S^{(1)}_\mathrm{EE}&=-T_7\int_{\rho_h}^{\rho_\epsilon}d\rho\int d\Sigma\, n^2\partial_n\sqrt{\gamma}
 \ - \ \frac{t_0}{32G}\left(V_{\mathbb{H}^3}-\frac{4}{3}\mu^2V_{S^2}\int_0^{u_\mathrm{max}} du \tanh^2u\right)~.
\end{align}
This already reproduces the massless result (\ref{eqn:S_EE-D7massless}), as it should since in the massless case the 
remaining variation of the brane Lagrangian did not contribute to the entropy.

The remaining thing is to calculate the contribution from the variation of the brane Lagrangian, i.e.\ the second term of (\ref{eqn:S-EE-D7}).
Implementing the integration bound $z(\rho,\tau,u)<\mu^{-1}$ is a bit tricky, since it links the three integration 
variables in a non-trivial way.
There is a nicer way to do the integral, which we give in App.~\ref{app:Poincare-z-integral}. 
Setting again $\cosh u=1/s$, the result reads
\begin{align}\label{eqn:massive-D7-bulk-integral}
 T_7\int_{z<\frac{1}{\mu}}d\rho\, d\Sigma\, n^2\partial_n\sqrt{\gamma}&=
 \frac{t_0}{48G}V_{S^2}\int_{a/l}^1ds\sqrt{1-s^2}\left[\mu^2\frac{1-s^2}{s}-\frac{1}{2}\mu^4s(1-2s^2)\right]~.
\end{align}
Combining that with (\ref{eqn:S-EE-massive1}), we thus find
\begin{align}
 \mathcal S^{(1)}_\mathrm{EE}&=-\frac{t_0 V_{S^2}}{32 G}\int_{a/l}^1ds\frac{\sqrt{1-s^2}}{s^3}\left(1-\frac{2}{3}\mu^2s^2(1+s^2)-\frac{1}{3}\mu^4s^4(1-2s^2)\right)~.
\end{align}
We note again that $\mu$ is proportional to $M_q \ell$ in the field theory, 
and this is indeed the dimensionless combination which we expect to appear
in the entanglement entropy.
To isolate the divergent and finite parts we expand the result for small $\epsilon$, 
which yields
\begin{align}\label{eqn:SEE-massive-D7-GGE}
 \mathcal S^{(1)}_\mathrm{EE}&=-\frac{t_0 V_{S^2}}{32 G} \left(\frac{l^2}{2 \epsilon ^2}+\frac{1}{6} \left(4 \mu ^2+3\right) \log
   \left(\frac{\epsilon }{2 l}\right)-\frac{1}{4}+\frac{4 \mu ^2}{9}-\frac{\mu ^4}{45}\right)~.
\end{align}
In the case of a CFT we had noted already below (\ref{eqn:bulk-gravity-EE-expanded}) that the regularization-scheme 
and state independent information is in the coefficient of the $\log$-term.
For a CFT deformed by a relevant deformation the analogous question has been studied 
from the field-theory side in \cite{Hertzberg:2010uv,Hertzberg:2012mn} and holographically in \cite{Hung:2011ta,Lewkowycz:2012qr}.
It turns out that the universal information is in the coefficients of the terms $\mu^{d-2-2n}\log(\mu\epsilon)$
for $2n\leq d-2$.
To match our result to that notation, we would expand $\log(\epsilon/(2\ell))=\log(\mu\epsilon)-\log(2\mu\ell)$.
The conclusion then is that both parts of the coefficient of the $\log$-term in (\ref{eqn:SEE-massive-D7-GGE})
are universal.
Comparing our result to the calculation in \cite{Kontoudi:2013rla}, we indeed find that the coefficients of the $\log$ terms agree.
Matching the finite and power-divergent terms is difficult, since they are regularization-scheme and state dependent.
While the regularization procedure is rather transparent in the gravitational entropy calculation, there are 
subtleties when the backreacted geometry is used.
We will come back to this issue below, after giving another calculation of the massive flavor entanglement entropy
where we keep track of these subtleties.

\section{Flavor entanglement entropy from the minimal area}\label{sec:double-integral}
We have already seen that the universal terms in the entanglement entropy due to massive flavors calculated by the 
gravitational entropy method agree with those found in \cite{Kontoudi:2013rla}.
To get a better understanding of the remaining terms we now compare to a computation with the double-integral formula 
proposed in \cite{Chang:2013mca}.
The basic idea here is to calculate the change in the area of the minimal surface yielding the entanglement entropy
in an efficient way.
This would usually involve calculating the linearized backreaction and evaluating
\begin{align}\label{eqn:double-integral-simplified}
 \mathcal S^{(1)}_\mathrm{EE}=\frac{1}{4G_\text{N}}\int_x \frac{1}{2}T_\text{min}^{\mu\nu}(x) \delta g_{\mu\nu}(x)~,
\end{align}
where $T_\text{min}^{\mu\nu}$ is the energy-momentum tensor corresponding to the minimal surface 
(up to coefficients the variation of the induced volume form on the minimal surface with respect to
the spacetime metric).
The linear backreaction can be calculated in terms of the probe brane data using the gravitational Green's function $G$,
resulting in the double-integral formula
\begin{align}\label{eqn:deltaEE-10d}
 \mathcal S^{(1)}_\mathrm{EE}=\frac{1}{4G_\text{N}} \int_{x, y}\frac{1}{2}T_\text{min}^{\mu\nu}(x)\,\frac{\kappa}{2}G_{\mu\nu,\rho\sigma}(x,y)T_\text{probe}^{\rho\sigma}(y)~.
\end{align}
The fact that the minimal surface is always of codimension $2$ allows for a crucial simplification: 
The details of the brane embedding in the internal space become largely irrelevant, and can be subsumed into an effective energy-momentum tensor for the
probe branes.
The formula (\ref{eqn:deltaEE-10d}) can then be reduced to 
\begin{align}\label{eqn:deltaEE}
 \mathcal S^{(1)}_\mathrm{EE}=\pi \int_{x_s, y_s}T_\text{min}^{\mu_s\nu_s}(x_s)G_{\mu_s\nu_s,\rho_s\sigma_s}(x_s,y_s)T_\text{eff}^{\rho_s\sigma_s}(y_s)~,
\end{align}
where the subscript $s$ refers to the fact that only the spacetime components 
(those corresponding to the non-compact part of the geometry) are summed over, and not those in the internal space.
Likewise, also the integral is only over the spacetime coordinates.
The details of the derivation can be found in \cite{Chang:2013mca}.

\subsection{Linearized backreaction}
Once we have seen that only the effective spacetime part of the brane energy-momentum tensor is relevant in (\ref{eqn:deltaEE}),
we can actually just as well calculate the backreaction of that source on the spacetime part of the metric 
and go back to (\ref{eqn:double-integral-simplified}) to get the change in the entanglement entropy.
We thus start by calculating the linearized backreaction of the D7-branes embedded via (\ref{eqn:D7-embedding}) into the AdS$_5\times$S$^5$ background.
For the AdS$_5$ part we use standard Poincar\'{e} coordinates and for the S$^5$ part (\ref{eqn:metric-S5}).
As discussed above, in the gravitational entropy calculation the radius of the spherical region $A$ entered the choice of coordinates,
and the use of the bulk geometry with $\ell\,{=}\,1$ consequently implied that $\ell$ entered the identification of bulk and boundary quantities.
This is not the case here, and the separation of the branes directly corresponds to the flavor mass.
To avoid confusion we replace (\ref{eqn:D7-embedding}) by $\cos\psi=m z$ for this section, and note that $M_q=\frac{\sqrt{\lambda}}{2\pi}m$.
With that embedding the induced metric on the brane reads
\begin{align}\label{eqn:D7inducedmetricPoincare}
 \gamma&=\frac{L^2}{z^2}\left(\frac{dz\otimes dz}{1-m^2z^2}+dx_\mu \otimes dx^\mu\right)+L^2(1-m^2z^2)g_{S^3}~,
\end{align}
and we refer to the first term as $\gamma_s$ and to the second as $\gamma_i$.
To get the effective energy-momentum tensor of (\ref{eqn:deltaEE}), we integrate out the internal part of
the D7-brane action
\begin{align}
 S_\text{D7-brane}=-T_7\int d^8y\sqrt{\gamma}
 =
 -\int d^5y_s \left(T_7\int d\Omega_3\sqrt{\gamma_i}\right)\sqrt{\gamma_s}=:-\int d^5y_s T_{5d}\sqrt{\gamma_s}~.
\end{align}
Since the brane direction $z$ parametrizes both, a spacetime and an internal direction, 
$\gamma_s$ still carries information on the embedding into the internal space and is {\it not} induced 
from the AdS$_5$ part of the bulk metric.
With $\sqrt{\gamma_i}=\sqrt{g_{\mathrm{S}^3}}L^3\sin^3\psi(z)$ we find
\begin{align}
 T_{5d}=T_7 V_{\mathrm{S}^3}L^3(1-m^2z^2)^{3/2}~.
\end{align}
From the AdS$_5$ perspective, we get a spacetime-filling brane with a position-dependent effective tension.
Coupling this brane to the effective five-dimensional bulk Einstein-Hilbert action (\ref{eqn:effective-5d-gravity-action}), the resulting
energy-momentum tensor appearing on the right hand side of Einstein's equations is
\begin{align}
 T_\text{eff}^{\mu_s\nu_s}&=-\frac{2}{\sqrt{g_s}}\frac{\delta S_\text{D7-brane}}{\delta g_{s\,\mu_s\nu_s}} =
 \frac{\sqrt{\gamma_s}}{\sqrt{g_s}}T_{5d}\gamma^{\mu_s\nu_s}=T_0 (1-m^2z^2)\gamma^{\mu_s\nu_s}~.
\end{align}
Nicely enough, this properly rescaled $T_{eff}^{\mu_s\nu_s}$ is conserved from the 5d perspective.
We can thus calculate the backreaction in the 5d effective picture, as advocated above.
Noting that the source respects translations and rotational invariance along the $x^{\mu_0}$ directions, we make the ansatz
\begin{align}\label{eqn:deltag-ansatz1}
 \delta g_s=\frac{L^2}{z^2}\left(f(z)dz\otimes dz+h(z)dx^{\mu_0}\otimes dx_{\mu_0}\right)~.
\end{align}
The functions $f$ and $g$ can then be determined by perturbatively solving Einstein's equations.
This fixes $f$ in terms of $h$ by the relation
\begin{align}\label{eqn:Einstein-fh}
 f(z)&=-\frac{t_0}{12}\big(1-m^2z^2\big)^2-zh'(z)~.
\end{align}
The function $h$ itself is not further constrained by the Einstein equations, and represents the remaining gauge freedom to
make $\mathcal O(t_0)$ changes to the $z$ coordinate.
To not spoil the asymptotically-AdS form of the metric, both of $f$ and $h$ along with their derivatives should be finite 
for $z\rightarrow 0$, and the equation thus fixes the constant part of $f$.

What we have so far is the form of $\delta g_s$ in the region covered by the D7 branes, $z<1/m$, and we still have
to join it to the unperturbed solution for the region $z>1/m$, to which the branes do not extend.
The relevant junction conditions are that the induced metric and the extrinsic curvature on the hypersurface $z=1/m$ agree.
The first condition yields $h(1/m)=0$. 
The extrinsic curvatures $K_{\mu\nu}=\frac{1}{2}\left(\mathcal L_n h\right)_{\mu\nu}$ with $h_{\mu\nu}=g_{\mu\nu}-n_\mu n_\nu$ are
\begin{align}
 K^{z>m^{-1}}_{\mu_0\nu_0}&=-L^{-1}g_{\mu_0\nu_0}~,&
 K^{z<m^{-1}}_{\mu_0\nu_0}&=\frac{\left(z h'(z)-2 h(z)-2\right)}{2 \sqrt{f(z)+1}}L^{-1}g_{\mu_0\nu_0}~.
\end{align}
Demanding those two to be equal at $z=1/m$ and using that $h(1/m)=0$, we find the additional condition $h^\prime(1/m)=0$.
Up to these requirements, the choice of $h$ is not constrained.

\subsection{Entanglement entropy}\label{sec:double-integral-EE}
With the linearized backreaction at hand, we can now
calculate the entanglement entropy from (\ref{eqn:double-integral-simplified}).
To derive the energy-momentum tensor corresponding to the original minimal surface,
we switch to spherical coordinates on the spatial part of the slices transverse to the AdS radial direction,
such that
\begin{align}
 dx^{\mu_0} dx_{\mu_0}=dt^2+dr^2+r^2d\Omega_2^2~.
\end{align}
The minimal surface can then be parametrized by
 $z=\ell s$\,, $r=\ell\sqrt{1-s^2}$ and $\Omega_2=\Omega_2(\varphi_1,\varphi_2)$.
 The induced metric on the minimal surface then is $\mathbb{H}^3$ in the form
 \begin{align}\label{eqn:induced-metric-minimal-surface}
 \gamma^{}_\mathrm{min}&=\frac{L^2}{s^2}\left(\frac{ds\otimes ds}{1-s^2}+(1-s^2)\,g^{}_{\mathrm{S}^2}\right)~.
\end{align}
Of the energy-momentum tensor  $T_\mathrm{min}$ we only need the diagonal part, since it will be contracted 
with $\delta g_s$, which is diagonal.
From (\ref{eqn:induced-metric-minimal-surface}) we find 
\begin{align}\label{eqn:minimal-surface-stress-tensor}
 \diag(T_\mathrm{min}^{\mu\nu})=\frac{s^2}{L^2}\left(\ell^2(1-s^2),0,\ell^2 s^2,\frac{1}{1-s^2},\frac{1}{1-s^2}\csc^2(\varphi_1)\right)~.
\end{align}
The entropy correction due to the change in the minimal area, (\ref{eqn:double-integral-simplified}), 
can then be evaluated with (\ref{eqn:deltag-ansatz1}) and (\ref{eqn:minimal-surface-stress-tensor}), which yields
\begin{align}
 \mathcal S^{(1)}_\mathrm{EE}&=\frac{1}{4G}L^3V_{S^2}\int_{\epsilon/\ell}^1 ds \frac{\sqrt{1-s^2}}{2s^3}  \left[\left(s^2+2\right) h(l s)-\left(s^2-1\right) f(l s)\right]~.
\end{align}
Note that the cut-off $z>\epsilon$ corresponds to $s>\epsilon/l$.
After replacing $f$ by (\ref{eqn:Einstein-fh}), we can use partial integration to reduce the integral to
\begin{align}\label{eqn:S_EE-double-integral-partial}
 \mathcal S^{(1)}_\mathrm{EE}&=-\frac{1}{4G}\frac{t_0}{24}L^3V_{S^2}\int_{\epsilon/\ell}^1 ds\frac{(1-s^2)^{3/2}}{s^3}\left(1-(mls)^2\right)^2
 \ + \ \frac{L^3V_{S^2}}{8G}h(\epsilon)\frac{\ell^2}{\epsilon^2}\left[1-\frac{\epsilon^2}{\ell^2}\right]^{3/2}~.
\end{align}
The dependence on $h$ is thus reduced to its value at the cut-off surface $z\,{=}\,\epsilon$, or, more precisely,
to the first few terms of its Taylor expansion around $z\,{=}\,0$, since it multiplies an asymptotic series.
This reduction should be expected, since different choices for $h$ are related by gauge transformations.
One would usually expect $h$ to drop out entirely, that it does not is due to the fact that we are dealing with
an infinite area and the choice of $h$ affects the regularization.
The remaining integral in (\ref{eqn:S_EE-double-integral-partial}) can then be performed easily.
With  $\mu\,{=}\,m l$ we find for the expansion around $\epsilon=0$
\begin{align}\label{eqn:SEE-double-integral-final}
 \mathcal S^{(1)}_\mathrm{EE}&=-\frac{t_0L^3V_{\mathrm{S2}}}{32G}  \left(
                 \frac{\ell^2}{6\epsilon^2}
                +\frac{4 \mu ^2+3}{6} \log \frac{\epsilon }{2\ell}
                +\frac{1}{4}+\frac{8 \mu^2}{9}+\frac{\mu ^4}{15}\right)
                +\frac{L^3V_{\mathrm{S}^2}}{8G}h(\epsilon)\left(\frac{\ell^2}{\epsilon^2}-\frac{3}{2}\right)~.
\end{align}
The ambiguities due to the free choice of $h$ reflect the freedom in the perturbed dual field theory to adjust the regularization scheme:
As emphasized already in \cite{Chang:2013mca}, the identification of the geometric bulk cut-off 
$z_\epsilon\,{=}\,\epsilon$ with a field-theory cut-off $\Lambda\,{=}\,1/z_\epsilon$ generally is spoiled 
by the change in the geometry due to the backreaction.
This corresponds to an $\mathcal O(N_f/N)$ adjustment of the regularization procedure in the dual field theory.
In fact, we could have used even more general ans\"atze than (\ref{eqn:deltag-ansatz1}).
We see, however, that the universal coefficient of the $\log$-term agrees with the one found in
the gravitational entropy (\ref{eqn:SEE-massive-D7-GGE}).

\subsection{Comparison to gravitational entropy}
We now want to discuss in some more detail how the backreaction method relates to the calculation in terms of the gravitational entropy.
Comparing the scheme-dependent finite and power-divergent terms is notoriously difficult, as it requires a matching of the 
regularization schemes.
However, before discussing that issue there is another subtlety left to be taken care of.
In the calculation of the gravitational entropy, adding the flavor branes to the setup does not seem to alter the regularization scheme:
once the cut-off procedure is fixed in the unperturbed theory, it is not affected by the probe branes, unless we choose to explicitly change it.
In the backreaction approach, on the other hand, the change in the bulk metric does imply that the interpretation of the bulk cut-off in
the dual theory is affected.
We thus still have to isolate the entanglement entropy contribution of the flavors from that due to a change in the regularization scheme.
To this end we will now determine the precise meaning of the cut-off at $z\,{=}\,\epsilon$ in the perturbed CFT,
and then fix the ambiguities in (\ref{eqn:SEE-double-integral-final}) by demanding that the regularization
schemes in the perturbed and unperturbed CFTs are the same.
In the massless case the perturbed metric is still AdS, and we could transform it to Poincar\'{e} coordinates.
The coordinate transformation depends on $h$, and the cut-off at $z\,{=}\,\epsilon$ then corresponds, 
depending on $h$, to different cut-offs $z^\prime\,{=}\,\epsilon^\prime(\epsilon)$ in Poincar\'{e} coordinates.
The latter can then be identified with the cut-off in the perturbed CFT.
The massive case is less straightforward, since the backreacted metric is not AdS anymore.
To relate the bulk and boundary cut-offs we will thus follow the covariant procedure discussed in \cite{Bousso:2009dm}.
The basic idea is to determine the minimal length $\delta$ which can be resolved in the cut-off CFT as follows:
One starts with a causal diamond of a given maximal spatial extent on the boundary, 
and extends it to a causal wedge in the asymptotically-AdS bulk spacetime.
If the spatial extent of the boundary causal diamond is small enough, its extension into the bulk 
will be entirely hidden behind the cut-off surface. 
The marginal case, where the extension into the bulk just touches the cut-off surface, then sets 
the minimal length $\delta$ in the CFT, corresponding to the given cut-off in the bulk theory.
The identification for metrics of the form (\ref{eqn:deltag-ansatz1})
has been worked out in \cite{Heemskerk:2010hk}. 
For our $g_s+\delta g_s$ it evaluates to
\begin{align}
 \delta&=\int_0^\epsilon dz\sqrt{\frac{1+f(z)}{1+h(z)}}=\epsilon+\frac{1}{2}\int_0^\epsilon dz\left(f(z)-h(z)\right)+\mathcal O(t_0^2)~,
\end{align}
where we have dropped terms of higher order in the backreaction to get the second equality.
The second term gives the $\mathcal O(t_0)$ change of the CFT cut-off.
Using (\ref{eqn:Einstein-fh}) and integration by parts, we can eliminate $h$ in the integrand,
which produces a boundary term at $z\,{=}\,\epsilon$ and yields
\begin{align}\label{eqn:cut-off-perturbed-CFT}
 \delta&=\epsilon - \frac{\epsilon}{2}\left[h(\epsilon)+\frac{t_0}{12}\left(1-\frac{2}{3}m^2\epsilon^2+\frac{1}{5}m^4\epsilon^4\right)\right]+\mathcal O(t_0^2)~.
\end{align}
To get a consistent result for this new cut-off, we would have to take into account that, due to the changed cut-off, 
there is an additional contribution to the entanglement entropy at $\mathcal O(t_0)$ from the original minimal surface, 
as emphasized in \cite{Chang:2013mca}.
Namely, we would have to add the part of the original minimal surface bounded between $z\,{=}\,\epsilon$
and the surface $z\,{=}\,\delta$, representing the new CFT cut-off in the unperturbed bulk geometry.
The more convenient variant of course is to just choose $h$ such that the meaning of the cut-off in the CFT is not changed.
We thus have to demand the expression in square brackets in (\ref{eqn:cut-off-perturbed-CFT}) to vanish, which fixes $h(\epsilon)$.
With this choice of $h$ the regularization procedure is then unaffected by the flavors and we have isolated their contribution
in the entanglement entropy (\ref{eqn:SEE-double-integral-final}), which becomes
\begin{align}\label{eqn:SEE-double-integral-massive-matched-cutoff}
 \mathcal S^{(1)}_\mathrm{EE}&=-\frac{t_0L^3V_{\text{S2}}}{32G}  \left(\frac{\ell^2}{2\epsilon^2}
                +\frac{4 \mu ^2+3}{6} \log \frac{\epsilon }{2\ell}
                -\frac{1}{4}+\frac{2 \mu^2}{3}+\frac{\mu ^4}{15}\right)~.
\end{align}
Comparing to the gravitational entropy (\ref{eqn:SEE-massive-D7-GGE}), we now find that, in addition to
the universal $\log$-terms, also the scheme-dependent power-divergent terms agree.
The only remaining difference is in the $m$-dependent finite terms.
But this should not come as a surprise, given that we have not precisely matched the regularization procedures: 
For the regularization of the minimal area we have chosen a cut-off at constant $z$, and then fixed the backreaction such
that the meaning of that cut-off in the field theory is preserved.
For the gravitational entropy calculation, on the other hand, we had chosen a different cut-off,
shown in Fig.~\ref{fig:AdS-S1H3-2}.

To elaborate a bit further on this point, we recall the identification 
of the cut-off bulk theory with the cut-off CFT in the approaches to the holographic 
renormalization group in \cite{Heemskerk:2010hk,Faulkner:2010jy}, focusing on a bulk scalar field $\phi$.
One rewrites the full bulk partition function $\mathcal Z[\phi^{}_0]$ in terms of the partition functions
$\mathcal Z_\mathrm{IR}$ on the cut-off spacetime and $\mathcal Z_\mathrm{UV}$ on the remaining part as
\begin{align}
 \mathcal Z[\phi^{}_0]&= 
 \int \mathcal D\phi^{}_\epsilon\:\mathcal Z^{}_\text{IR}[\phi^{}_\epsilon]\,\mathcal Z^{}_\text{UV}[\phi^{}_0,\phi^{}_\epsilon]~,
\end{align}
where $\phi_0$ and $\phi_\epsilon$ are the boundary values on the conformal boundary and on the cut-off surface, respectively.
The correlators of the dual operator in the cut-off CFT are then obtained from
\begin{equation}\label{eqn:AdS/cutoffCFT}
 \mathcal Z^{}_\text{IR}[\phi^{}_\epsilon]
 =\Big\langle \exp\big\lbrace \int \phi^{}_\epsilon\mathcal O\big\rbrace\Big\rangle_{\text{CFT},\epsilon}~.
\end{equation}
In the semiclassical limit, where the bulk path integral is dominated by the on-shell action,
this gives the one-point function as the normal derivative of $\phi$ at the cut-off.
When the cut-off surface approaches the conformal boundary, this turns into the usual AdS/CFT prescription where the subleading
mode on the boundary gives the one-point function.
We now turn to the D7 branes.
For the slipping mode $\psi$, the boundary-dominant solution was chosen
in (\ref{eqn:D7-embedding}), which sources as dual operator the mass term of the flavor fields. 
There is no contribution from the subdominant mode, which would produce a vacuum expectation
value and corresponds to a deformation of the state.
This clear split is lost in the cut-off theory:
the slipping mode (\ref{eqn:D7-embedding}) does not satisfy a pure Dirichlet or Neumann boundary 
condition at the cut-off surface.
Evaluating (\ref{eqn:AdS/cutoffCFT}) thus produces a non-vanishing one-point function,
and adding the flavor branes perturbs the Hamiltonian and the state in the cut-off CFT.
As the cut-off approaches the conformal boundary, we get back to the pure deformation of the Hamiltonian
in the full CFT.
However, here we have studied the cut-off CFT, and
choosing different cut-off surfaces corresponds to different 
admixtures of perturbations to the state.
As pointed out in \cite{Hung:2011ta}, that affects the finite terms of the entropy:
while the divergent parts are generally independent of the state, i.e.\
the full density matrix from which the entanglement entropy is calculated,
this is not the case for the finite parts.
We thus conclude that the universal terms agree in both calculations, and even the scheme-dependent power-divergent
parts do, which is as much as we can expect.

\section{Conclusions}\label{sec:conclusion}

In this work we have studied the gravitational entropy introduced in \cite{Lewkowycz:2013nqa},
with a focus on practical applications in AdS/CFT.
While its conceptual relevance for the calculation of entanglement entropies in AdS/CFT is clear, as it permits an actual derivation of the 
minimal-area prescription \cite{Ryu:2006bv}, one may na\"ively expect it to be of little practical value, 
precisely because it reduces to the minimal area.
We have shown that it does offer practical advantages, too, focusing on the case where the bulk 
theory is perturbed by the addition of probe branes.
The minimal-area prescription does not allow to efficiently exploit the probe approximation, 
and one has to calculate the backreaction to get the leading correction to the CFT
entanglement entropy.
Our general discussion of probe branes in Sec.~\ref{sec:GGE-probe-branes} has shown that 
for the gravitational entropy, on the other hand, we can directly exploit that the branes 
perturb an on-shell configuration, and thus avoid calculating their backreaction.
The line of arguments is in fact not limited to the case of probe branes.
Rather, generally when a deformation of the gravity theory is considered only perturbatively,
it should be possible to use analogous arguments to get the entropy without calculating the backreaction.
This applies for the gravitational entropy itself, and in particular for the case where
it is used with AdS/CFT to calculate entanglement entropies for the dual CFT.

As a specific application we studied holographically the contribution of various types of
flavors in \N{4} SYM theory to the entanglement entropy of a spherical region.
We have validated our method by comparing to existing calculations 
in the literature.
In comparison to the computation of the backreacted minimal area, which in particular for the
case of massive flavors involves non-trivial techniques like a smearing of the flavor branes 
over the internal space, the calculation has become very straightforward.
For the case of massive flavors we also compared to another approach, which simplifies 
the calculation from a different perspective: 
the double-integral method of \cite{Chang:2013mca} does in fact boil down to computing the backreaction, 
but only of an effective source on the AdS part.
This allowed for an independent concise derivation, which confirmed our result obtained from the gravitational entropy.
After subtleties in the regularization procedure due to the backreaction were taken into account, also the scheme-dependent
power-divergent parts agreed.
It is worthwhile to compare the two methods in a bit more detail.
The double-integral formula offers a rather drastic simplification of the backreaction approach.
Its simplest form has limitations when the brane sources non-metric bulk fields which
already have background values in the bulk solution.
But if that is not the case, one just needs the backreaction of an effective source, with the effective 
tension obtained from the full brane action, on the non-compact part of the bulk geometry.
It thus allows to be agnostic, to some extent, about the details of the internal space.
The gravitational entropy method, on the other hand, avoids the complications of the backreaction altogether,
and only needs the brane action.
It also does not interfere with the interpretation of the bulk cut-off in the dual CFT, as the $n\,{=}\,1$
solution is not deformed.
It does, however, need the one-parameter family of bulk geometries with varying period of the S$^1$.
As we have seen for the brane embeddings, one does not need the full solutions, 
just the background geometry is enough.
Yet, finding that family of geometries can be non-trivial.
Depending on the case at hand, one of the approaches or the other may thus be more convenient.
The gravitational entropy method is particularly easy to implement once the one-parameter family of
bulk solutions is known.
It thus calls for further application, in particular for a spherical region $A$.
For another choice, where $A$ is a half space, the family of bulk solutions has been discussed in \cite{Lewkowycz:2013nqa}.
It should, among other things, be possible along the arguments given in Sec.~\ref{sec:GGE-probe-branes},
to incorporate non-trivial worldvolume gauge fields and study entanglement in the dual CFT 
at finite density.

\begin{acknowledgments}
We thank Han-Chih Chang, Kristan Jensen, Matthias Kaminski and Andy O'Bannon for useful discussions.
The work of AK is supported in part by the US Department of Energy under grant number DE-FG02-96ER40956.
CFU is supported by {\it Deutsche Forschungsgemeinschaft} through a research fellowship.
\end{acknowledgments}

\appendix

\section{Integrating \texorpdfstring{$\delta_n \sqrt{\gamma}$}{varied brane Lagrangian} for separated D7}\label{app:Poincare-z-integral}
For completeness we now explain in a bit more detail how the result (\ref{eqn:massive-D7-bulk-integral}) is obtained.
With (\ref{eqn:determinants}) the left hand side of (\ref{eqn:massive-D7-bulk-integral}) becomes
\begin{align}
 T_7\int_{z<\frac{1}{\mu}}d\rho\, d\Sigma\, \delta_n\sqrt{\gamma}&=
 T_7\,\mu^2\int_{z<\frac{1}{\mu}}d\rho d\tau du \int_{S^2\times S^3}\rho^3\sinh^2(u)(1-\mu^2z^2)
 \frac{(\tilde\rho^2\partial_\rho z)^2-(\partial_\tau z)^2}{3\rho^{2}\tilde\rho^{4}}~,
\end{align}
where $z$ again denotes $z(\rho,\tau,u)$ as given in (\ref{eqn:coord-transf}) and the volume forms on S$^2$ and S$^3$ are implicit.
Implementing the restriction to $z<1/\mu$ in our coordinates is a bit tricky, but we can actually circumvent it:
as an intermediate step, we perform the integration over the full AdS$_5$ corresponding to $z\in\mathbb{R}^+$,
and then subtract the part with $z>\mu^{-1}$.
Substituting $\cosh u=1/s$ we find 
\begin{align}\label{eqn:bulk-integral-split1}
\begin{split}
  T_7\int_{z<\frac{1}{\mu}}d\rho d\Sigma\,\delta_n\sqrt{\gamma}&=
 \frac{t_0V_{S^2}}{48G}\int_{a/l}^1ds\sqrt{1-s^2}\left[\mu^2\frac{1-s^2}{s}-\frac{1}{2}\mu^4s(1-2s^2)\right]\\
 &\hphantom{=}-T_0\,\mu^2V_{S^2}\int_{z>\frac{1}{\mu}}d\rho d\tau du \rho^3\sinh^2(u)(1-\mu^2z^2)
 \frac{(\tilde\rho^2\partial_\rho z)^2-(\partial_\tau z)^2}{3\rho^{2}\tilde\rho^{4}}~.
\end{split}
\end{align}
Since the integral is over the  $n\,{=}\,1$ geometry, which is just Euclidean AdS$_5$, 
we might as well switch to Poincar\'{e} coordinates for the remaining part.
We could try to invert the coordinate transformation (\ref{eqn:coord-transf}), 
but we find it is easier to use explicit parametrizations of the AdS hyperboloid. 
For $L=1$ the Euclidean versions of the parametrizations in \cite{Casini:2011kv} read
\begin{align}
 y_{-1}&=\rho\cosh u~, & y_0&=\tilde\rho\sin\tau~, & y_d&=\tilde\rho\cos\tau~, & y_a=\rho\sinh u \ \omega_a~,
\end{align}
for the hyperbolic slicing, where $a=1,..,d-1$ and $\sum\omega_a^2=1$. For the Poincar\'{e} coordinates we use
\begin{align}\label{eqn:Poincare-parametrization}
 y_{-1}+y_d&=z^{-1}~, & y_{-1}-y_d&=z+z^{-1}x^\mu x_\mu~, & y_i&=x^i/z~,
\end{align}
where $i=0,..,d-1$. Using these parametrizations and the explicit expression for $z(\rho,\tau,u)$,
we can then express the terms of the integrand in (\ref{eqn:bulk-integral-split1}) as
\begin{align}\label{eqn:integrand-poincare}
 \partial_\tau z&=t z~, &
 \tilde\rho^2\partial_\rho z&=z^2y_{-1}/\rho-\rho z~,&
 \tilde\rho^2&=t^2z^{-2}+y_d^2~.
\end{align}
The volume form transforms to $d\rho d\tau du \rho^3\sinh^2(u)=dz dr dt\,r^2z^{-5}$.
While the bound $z>1/\mu$ is implemented straightforwardly in Poincar\'{e} coordinates, 
the integrand itself becomes rather bulky.
However, for $z>1$ we see from (\ref{eqn:Poincare-parametrization}) and (\ref{eqn:integrand-poincare}) that $\tilde\rho>0$. 
The integrand thus has no poles and the integral can be performed straightforwardly.
The second line of (\ref{eqn:bulk-integral-split1}) then evaluates to zero after the $t$ and $r$ integrations,
\begin{align}
 -T_0\,\mu^2V_{S^2}\int_{z>\frac{1}{\mu}}dz dr dt\, r^2 z^{-5}(1-\mu^2z^2)
 \frac{z^2\left((\rho-zy_{-1}/\rho)^2-t^2\right)}{3\rho^2\tilde\rho^4}&=0~.
\end{align}
We have also confirmed this result numerically, and 
the first line of (\ref{eqn:bulk-integral-split1}) thus is the final result.

\bibliography{flavorbranes.bib}
\end{document}